\documentclass[aps,twocolumn,showpacs,superscriptaddress]{revtex4}
\usepackage{bbm}
\usepackage{mathrsfs}
\usepackage{epsfig}
\usepackage{graphicx}
\usepackage{amsfonts}
\usepackage[figuresright]{rotating}
\usepackage{amssymb}
\usepackage{amsmath}
\usepackage{dcolumn}
\usepackage{bm}

\def\avg#1{\langle#1\rangle}

\def\be{\begin{equation}} \def\ee{\end{equation}}
\def\bea{\begin{eqnarray}} \def\eea{\end{eqnarray}}

\def\nn{\nonumber}
\def\pp{\parallel}

\newcommand{\ket}[1]{| #1 \rangle}
\newcommand{\bra}[1]{\langle #1 |}

\begin{document}
\title{Sign problem free quantum Monte-Carlo study 
on thermodynamic properties and magnetic phase 
transitions in orbital-active itinerant ferromagnets} 
\author{Shenglong Xu}
\affiliation{Department of Physics, University of California,
San Diego, California 92093, USA}
\author{Yi Li}
\affiliation{Princeton Center for Theoretical Science, Princeton
University, Princeton, New Jersey 08544, USA}
\author{Congjun Wu}
\affiliation{Department of Physics, University of California,
San Diego, California 92093, USA}

\begin{abstract}
The microscopic mechanism of itinerant ferromagnetism is a long-standing 
problem due to the lack of non-perturbative methods to handle strong 
magnetic fluctuations of itinerant electrons.
We have non-pertubatively studied thermodynamic properties and magnetic 
phase transitions of a two-dimensional multi-orbital Hubbard model 
exhibiting ferromagnetic ground states.
Quantum Monte-Carlo simulations are employed, which are
proved in a wide density region free of the sign problem 
usually suffered by simulations for fermions.
Both Hund's coupling and electron itinerancy are essential for 
establishing the ferromagnetic coherence.
No local magnetic moments exist in the system as a priori, 
nevertheless, the spin channel remains incoherent showing the 
Curie-Weiss type spin magnetic susceptibility down to very low temperatures 
at which the charge channel is already coherent exhibiting a weakly 
temperature-dependent compressibility.
For the SU(2) invariant systems, the spin susceptibility further grows 
exponentially as approaching zero temperature in two dimensions.
In the paramagnetic phase close to the Curie temperature, 
the momentum space Fermi distributions exhibit strong
resemblance to those in the fully polarized state. 
The long-range ferromagnetic ordering appears when the symmetry is
reduced to the Ising class, and the Curie temperature is accurately
determined. 
These simulations provide helpful guidance to searching for
novel ferromagnetic materials in both strongly correlated 
$d$-orbital transition metal oxide layers and the $p$-orbital 
ultra-cold atom optical lattice systems.
\end{abstract}
\pacs{Subject area: magnetism, condensed matter physics,
atomic molecular and optical physics}
\maketitle

\section{Introduction}
Itinerant ferromagnetism (FM) is one of the central topics of condensed
matter physics \cite{stoner1938,zener1951,slater1953a,
anderson1955,lieb1962,nagaoka1966,herring1966,hertz1976,
murata1972,moriya1985,hirsch1989,mielke1993a,tasaki1992,millis1993,
baberschke2001,belitz2005,jackiewicz2005,lohneysen2007,maslov2009,chen2013,
li2014,sang2014,mattis2006}.
It  has also become a research focus
both experimental and theoretical of ultra-cold atom physics  
\cite{duine2005,Jo2009,zhangSZ2010,berdnikov2009,
pekker2011a,chang2010,cui2014,pilati2014}.
The mechanism of itinerant FM has been a long-standing problem.
Stoner proposed the exchange interaction among electrons 
with parallel spins as the driving force for itinerant FM \cite{stoner1938}.
Along this direction, the local density approximation (LDA)
of the density functional theory has achieved great success
\cite{kohn1965,barth1972}.
For example, the ground state magnetic moments of FM metals can be
calculated accurately \cite{moruzzi1978}.
The implementation of correlation effects in LDA has also been
improved by the methods of LDA+U \cite{himmetoglu2013},
LDA+DMFT(dynamical mean-field theory) \cite{georges1996,gull2011,held2001},
and LDA+GP (Gutzwiller projection)
\cite{gutzwiller1963,zhuang2009,weber2001}.

Nevertheless, itinerant FM systems are also  strongly correlated,
and their physics is often non-perturbative. 
Usually repulsive interactions need to be sufficiently strong 
to overcome the kinetic energy cost of polarizing electron spins,
and thus itinerant FM has no well-controlled weak-coupling starting point. 
The Stoner criterion overlooks correlation effects
among electrons with opposite spins \cite{mattis2006}:
Electrons can delicately organize their wavefunctions
to reduce repulsions and still remain unpolarized even in the presence of
strong interactions.
For example, the Lieb-Mattis theorem proves that the ground state
of a rigorously one-dimensional (1D)
system is a spin singlet no matter how strong the
interaction is \cite{lieb1962}.

It is more appropriate to start with electron orbitals to construct
lattice model Hamiltonians to address the strong correlation aspect 
of itinerant FM.
Exact theorems establishing FM, which are usually based on lattice 
models, are indispensable to provide reference 
points for further investigations.
Well-known examples include the Nagaoka theorem \cite{nagaoka1966,
shastry1990,tian1991,tasaki1989,liu2012,gu2014}, which applies
to the infinite $U$ Hubbard models in
two and above dimensions with doping a single hole on the half-filled
background, and the ``flat-band'' FM in certain lattices
with dispersionless band structures \cite{mielke1991,tasaki1992}.
In the former case, FM arises because the  
spin polarized background maximally
facilitates the hole's coherent hopping, while in the latter case, the band
flatness reduces the kinetic energy cost for polarizing spin to zero.

One central issue of itinerant FM is the role of orbital degeneracy 
which widely exists in FM metals.
Hund's coupling is a prominent feature in multi-orbital systems, 
which favors electrons on the same site to align their spins.
However, Hund's coupling is local physics which usually cannot 
polarize itinerant electrons in the absence of local moments.
Under what precise conditions Hund's coupling can lead to the global FM 
coherence in itinerant systems is still an open question.
 
The difficulty in achieving unambiguous FM ground states is only 
one side of the story of strongly-correlated itinerant FM, the
finite-temperature thermodynamic properties  are another challenge 
\cite{moriya1973,moriya1985,stoehr2006,fazekas1999}.
At first looking, it might not look so striking:
the ferromagnetic susceptibilities show 
the standard mean-field Curie-Weiss (CW) law in the off-critical region as
\bea
\chi(T)=\frac{C}{T-T_0},
\label{eq:curie}
\eea
where $C$ is the Curie constant \cite{rhodes1962} and $T_0$
is the Curie temperature at the mean-field level.
The CW law manifests spin incoherence, which is common in 
the paramagnetic state based on local moments.
But it is difficult to understand in itinerant FM metals
still possessing Fermi surfaces.
For example,  the transport and the charge channel 
properties, such as resistance and compressibility, remain
metallic, {\it i.e.}, they are featured by the Fermi surface physics. 

Within the itinerant picture,
the Pauli magnetic susceptibility augmented by the random phase approximation 
(RPA) yields $\chi(T)
\propto 1/(T^2-T_0^2)$, but it is not commonly observed in experiments
\cite{moriya1985,stoehr2006,fazekas1999}.
In fact, the CW law in FM metals applies to a wide range of temperatures 
$T_f\gg T > T_0$ ($T_f$ is the Fermi temperature)
which shows spin-incoherence well below $T_f$.
The reason is that RPA treats the paramagnetic phase 
as a weakly correlated Fermi liquid state with slightly thermally broadened 
Fermi distributions.
Actually, this phase is rather complicated: Dynamic FM domains strongly
fluctuate, which is beyond the RPA description and is difficult to 
handle analytically. 
The paramagnetic state of itinerant FM exhibits much higher entropy 
capacity than the usual weakly correlated paramagnetic Fermi liquid state, 
which significantly suppresses the genuine Curie 
temperature $T_c$, or, the renormalized one, away from 
the mean-field value $T_0$.
Consequently, $T_c$ is often significantly overestimated by weak
coupling approaches \cite{moriya1985,stoehr2006,fazekas1999}.

A key question is how  itinerant systems can
exhibit the CW law and further develop FM purely based on 
itinerant electrons without involving local moments such that 
the charge channel remains coherent?
Significant efforts have been made, including the self-consistent 
renormalization theory including spin mode coupling 
\cite{murata1972,moriya1973,moriya1985}, the direct exchange from 
the Coulomb integral \cite{hirsch1989,hirsch1989a}, spin incoherence 
due to Hund's coupling \cite{aron2014}, and the orbital-selective 
Mott transition \cite{demedici2009,anisimov2002}.
An important progress is that the CW law can be obtained from the 
combined method of LDA+DMFT \cite{lichtenstein2001} away from 
the critical region.
However, none of these methods are non-perturbative in nature.

Another issue is the nature of the FM phase transitions
in FM metals, which has been been widely studied but is still 
under intensive debates \cite{hertz1976,millis1993,belitz2005,
lohneysen2007,maslov2009,jackiewicz2005}.
Compared to the superconducting phase transitions
in which the fermion degree of freedom is gapped below transition
temperatures, the FM phase transitions are more involved
because systems remain gapless across transitions 
due to the existence of Fermi surfaces. 
The FM domain fluctuations combined with the Landau damping
of particle-hole excitations around Fermi surfaces complicates 
FM transitions.
It would be important to perform a non-perturbative study.


Recently,  the ground states of a multi-orbital Hubbard model
have been proved fully spin polarized in the strong coupling 
regime in the 2D square and 3D cubic lattices \cite{li2014}
by two of the authors and Lieb.
It is showed that inter-orbital Hund's coupling combined with 
electron itinerancy in the quasi-1D band structure drive the FM ground states.
Compared to the Nagaoka FM, this new theorem proves a stable FM phase 
with nodeless ground state wavefunctions over the entire
electron density region $0<n<2$, where $n$ is the occupation number
per site, thus it sets up a solid starting point for further 
studying the strong correlation aspect of itinerant FM.
It also opens up the possibility of performing sign-problem
free quantum Monte Carlo (QMC) simulations away from half-filling
by employing the bases under which the many-body Hamiltonians
satisfy the Perron-Frobenius condition.

Although this theorem only sets up the ground state properties, 
it establishes an unambiguous FM phase as a starting point for 
further studying both thermodynamic properties and magnetic phase 
transitions over a wide region of electron density.
In order to handle the strong magnetic fluctuations, QMC simulations
would be the ideal method, however, they usually suffer the notorious 
sign problem for fermions and thus are generally speaking
inapplicable for itinerant FM.
Remarkably, we prove that for the systems in which the ground
state FM theorem mentioned above \cite{li2014} applies, the fermion
sign problem can be eliminated in the entire electron density region.
This provides a new opportunity to study the finite temperature 
thermodynamic properties and magnetic phase transitions in an 
asymptotically exact way.


For later convenience, we briefly discuss the FM critical fluctuations
which are particularly important in two-dimensions.
According to the Landau-Ginzburg-Wilson paradigm of critical phenomena,
$T_c$ is suppressed from $T_0$ but remains finite in 3D.
As $T$ is lowered from $T_0$ and approaches $T_c$, the system crosses 
over from the mean-field region to the critical region, and 
 $\chi(T)\propto (T-T_c)^{-\gamma}$ 
due to non-Gaussian fluctuations and $\gamma$ is the critical exponent.
In 2D, $T_c$ remains finite if the system symmetry is reduced 
to the Ising class, or, the easy axis class.
However, for the isotropic class, thermal fluctuations
suppress $T_c$ to zero according to the Mermin-Wagner theorem 
\cite{mermin1966,zaleski2008}.
Nevertheless, even in this case the mean-field $T_0$ is still an 
important temperature scale below which the FM order 
develops its magnitude.
However, the orientation fluctuations of the FM order 
suppress the long-rang-order.
In other words, this region is characterized by fluctuating
FM domains and the correlation length increases
exponentially as lowering temperatures. 
Consequently, the FM susceptibility  deviates
from the CW law and crosses over into an exponential growth. 

In this article, we will present a systematic non-perturbative study on
thermodynamic properties and magnetic phase transitions
of itinerant FM by performing the sign-problem 
free QMC simulations.
Our results show that itinerant FM can indeed exhibit both spin 
incoherence and charge coherence simultaneously without forming 
local moments.
In other words, the system exhibits the feature of the CW 
metal as a combined effect of Hund's coupling and electron itinerancy.
The model we simulate can be realized in both $d$-orbital transition
metal oxide layer and $p$-orbital ultra-cold atom optical lattices,
which do not contain local moments as a priori.
The spin magnetic susceptibility exhibits the CW law 
as a signature of spin incoherence, while, the compressibility
weakly depends on temperature as a consequence of itinerancy.
The mean-field Curie temperature $T_0$ is extracted based on the CW
law in the off-critical region,
which is much lower than the temperature scale of charge coherence  
$T_{ch}$.
The filling dependence of $T_0$ is calculated and the
maximal $T_0$ reaches one tenth of the hopping integral. 
The Fermi distribution functions in momentum space are calculated
in the strongly correlated paramagnetic phase.
The fermion occupation numbers are strongly suppressed from 
the saturated value even 
for wavevectors close to the center of the Brillouin zone.
When entering the critical region, for the SU(2) symmetric models,
$\chi(T)$ grows exponentially.
The true FM long range order is achieved by reducing the
model symmetry to the Ising class and the FM critical 
temperature $T_c$ is determined accurately by the finite size
and critical scaling.

The rest part of this article is organized as follows.
In Sect. \ref{sect:model},
the model Hamiltonian is introduced, and the QMC method for
this model is proved free of the sign problem.
The QMC simulations on the thermodynamic properties
in the off-critical region is presented in Sect. \ref{sect:off},
and the momentum space Fermi distributions are calculated
in Sect. \ref{sect:fermi}.
The results in the critical region are
presented in Sect. \ref{sect:critical}.
In Sect. \ref{sect:discussion}, we discuss the physics when the
conditions for the absence of the sign problem are loosed.
In particular, the simulations in the presence of a small inter-chain 
hopping, in which the sign problem appears but is not severe,
are presented.
The experimental realizations are discussed in Sect. \ref{sect:exp}.
Conclusions are made in Sect. \ref{sect:conc}.

\section{The Model Hamiltonian and the absence of the sign problem}
\label{sect:model}
In this section, we present the model Hamiltonians, whose ground
states were proved to be ferromagnetic \cite{li2014}.
Furthermore, we also explain that the QMC sign-problem is absent, 
and thus, this model provides
an ideal preliminary to study the thermodynamic properties and magnetic 
phase transitions of strongly-correlated itinerant FM in a controllable way.

\subsection{The model Hamiltonians}
We consider the case of the 2D square lattice:
On each site there are two orthogonal orbitals forming a quasi-1D
band structure.
For simplicity, below we use the 2D $p$-orbital system as an example,
and the physics is also valid for the $d_{xz}$ and $d_{yz}$-orbitals
systems in 2D.
The relevance of this model to the current experiment efforts of 
searching for novel itinerant FM systems will be
discussed in Sect. \ref{sect:discussion}.
For the band structure, we only keep the $\sigma$-bonding $t_\pp$
term, i.e., electrons in the $p_{x(y)}$-orbital only move longitudinally along the
$x(y)$-direction, respectively.
The following Hamiltonian is defined in the square lattice as
\bea
H_{kin,\parallel}&=&-t_\pp \sum_{\vec r, \sigma} \Big \{ p_{x\sigma}^\dagger(\vec r+\hat e_x)
p_{x\sigma}(\vec r)
+p_{y\sigma}^\dagger(\vec r +\hat e_y)\nn \\
&\times& p_{y\sigma}(\vec r )
+ h.c. \Big \} -\mu \sum_{\vec r} n(\vec r),
\label{eq:kin}
\eea
in which we neglect the small transverse bonding $t_\perp$-term.
For realistic $p$-orbital systems, the sign of $t_\pp$ is negative
due to the odd parity of $p$-orbital Wannier wavefunctions.
Nevertheless, for the bipartite lattice such as the square lattice,
the sign of $t_\pp$ can be flipped by a gauge transformation.
Without loss of generality, $t_\pp$ is scaled to 1 below,
which serves as the unit for all other quantities
carrying energy unit in this article. 

The interaction part $H_{int}$ contains the standard
multi-orbital Hubbard interaction  \cite{roth1966,kugel1973,
cyrot1975,oles1983} as
\bea
H_{int}&=& U\sum\limits_{\vec r,a=x,y}
n_{a,\uparrow}(\vec r)n_{a,\downarrow}(\vec r)
+V\sum_{\vec r} n_x(\vec r) n_y(\vec r) \nn \\
&-&J \sum_{\vec r} \Big\{ \vec{S}_x(\vec r)\cdot
\vec{S_y}(\vec r)-\frac{1}{4}n_x(\vec r) n_y(\vec r) \Big\}
\nn \\
&+&\Delta\sum_{\vec r}
\Big\{p_{x\uparrow}^\dagger(\vec r) p_{x\downarrow}^\dagger (\vec r)
p_{y\downarrow} (\vec r) p_{y\uparrow} (\vec r)
+h.c.\Big\},
\label{eq:int}
\eea
where $a=x,y$ referring to the orbital index;
$n_{a,\sigma}=p_{a,\sigma}^\dagger p_{a,\sigma}$ and
$n_a=n_{a,\uparrow}+n_{a,\downarrow}$;
$\vec{S}_{a}=p_{a,\alpha}^\dagger \vec \sigma_{\alpha\beta}p_{a,\beta}$
is the spin operator of the $a$-orbital.
The $U$ and $V$-terms describe the intra- and inter-orbital Hubbard
interactions, respectively; the $J$-term is Hund's coupling and
$J>0$ represents its FM nature; the $\Delta$-term describes
the pairing hopping process between two orthogonal orbitals.

In order to gain an intuitive understanding of the interaction parameters,
let us consider a single site problem.
There are in total six states which can be classified as
a set of spin triplet states and three different spin singlet
states.
The triplet states are with energy $V$, defined as
\bea
p^\dagger_{x,\uparrow} p^\dagger_{y,\uparrow}|0\rangle, ~~
\frac{1}{\sqrt 2}(p^\dagger_{x,\uparrow} p^\dagger_{y,\downarrow}
+p^\dagger_{x,\downarrow} p^\dagger_{y,\uparrow})|0\rangle, ~~
p^\dagger_{x,\downarrow} p^\dagger_{y,\downarrow}|0\rangle,
\eea
respectively,
where $|0\rangle$ is the vacuum state.
The other three spin singlet states are
\bea
\frac{1}{\sqrt 2}(p^\dagger_{x,\uparrow} p^\dagger_{y,\downarrow}
-p^\dagger_{x,\downarrow} p^\dagger_{y,\uparrow})|0\rangle, ~~
p^\dagger_{x,\uparrow} p^\dagger_{x,\downarrow}|0\rangle,~~
p^\dagger_{y,\uparrow} p^\dagger_{y,\downarrow}|0\rangle,
\eea
among which the first one involves both orbitals and its
energy is $V+J$; the other two singlets only occupy
the same orbital with the average energy $U$
and the hybridization matrix element
between them is $\Delta$.
In the limit of $U\rightarrow +\infty$, the states of
$p^\dagger_{x,\uparrow} p^\dagger_{x,\downarrow}|0\rangle$,
and $p^\dagger_{y,\uparrow} p^\dagger_{y,\downarrow}|0\rangle$ are
projected out.
Nevertheless, the other four doubly occupied states
are kept in the physical Fock space, including one set of spin-triplet
states and one inter-orbital singlet state.

The ground states of the Hamiltonians Eq. \ref{eq:kin} plus Eq. 
\ref{eq:int} are fully spin polarized at any generic filling $n$
for arbitrary values of $V$  under the condition
that  $U\rightarrow +\infty$ and $J>0$.
The detailed proof and its generalization to the 3D cubic lattice
are presented in Ref. [\onlinecite{li2014}].
Below we present an intuitive physical picture.
The band structure of Eq. \ref{eq:kin} is quasi-1D, consisting
of orthogonal rows and columns, and electrons do not transit among
different lines.
In the absence of Hund's coupling, then the intra-chain physics 
in the limit of $U\rightarrow +\infty$ would correspond to the 
1D infinite-$U$ Hubbard model whose ground states are highly 
degenerate regardless of the spin configurations.
Then let us turn on $J>0$, and the inter-chain Hund's coupling 
lifts the degeneracy and selects the fully polarized state as 
the unique ground state: When one electron in a row meets another
one in a column at the crossing site, their spins are aligned
to save the energy of $J$.
Thus different from the usual case that Hund's coupling can only 
polarize electrons on the same site. 
Remarkably, in this case it does polarize electrons in the entire
system 
\cite{li2014,chen2013}.
Although the electron band structure is quasi-1D, interactions
couple electron spins in different chains together, and thus,
the FM correlations and ordering are genuinely 2D, or, 3D. 

For completeness, we also present the Hamiltonian of the inter-chain
hopping with a small value of $t_\perp$ as
\bea
H_{kin,\perp}&=&-t_\perp \sum_{\vec r, \sigma} 
\Big \{ p_{x\sigma}^\dagger(\vec r+\hat e_y)
p_{x\sigma}(\vec r) \nn \\
&+&p_{y\sigma}^\dagger(\vec r +\hat e_x) p_{y\sigma}(\vec r ) 
+ h.c. \Big \},
\label{eq:Hperp}
\eea
which will be used in Sect. \ref{sect:tperp}.
Again, in the square lattice the sign of $t_\perp$ can be flipped
by a gauge transformation, and without loss of generality, it is assumed
to be positive. 
We set $t_\perp=0$ in most part of this article except in 
Sect. \ref{sect:tperp}.

\subsection{The absence of the QMC sign problem}
\label{sect:sign}

The many-body Hamiltonian matrix of Eq. \ref{eq:kin}
plus Eq. \ref{eq:int} possesses an important sign structure
in the limit of $U\rightarrow +\infty$ under which the ground state
FM theorem applies \cite{li2014}.
In the coordinate representation, a convenient set of many-body bases
are defined by ordering fermions according to their real space positions
along one row by another and then along one column by another.
The periodical and anti-periodical boundary conditions are
employed for each chain if the particle number in that chain is
odd and even, respectively, which is feasible because
the particle number in each chain is separately conserved.
This particular choice of boundary conditions should not change the bulk
physics.
Under these bases and boundary conditions, in the limit of
$U\rightarrow +\infty$, the electron hopping term and the spin-flip
term from Hund's coupling do not change the sequence of fermion ordering.
When electrons hop across the boundary, no extra minus
sign appears either due to the above boundary condition.
Then the many-body Hamiltonian matrix
satisfies the prerequisite of the Perron-Frobenius theorem: All
the non-zero off-diagonal matrix elements are either $-t$ or
$-J$ arising from the kinetic energy term and Hund's coupling,
respectively, and thus they are semi-negative-definite.
We do not need to consider the pair hopping process which
is completely suppressed in the limit of $U\rightarrow +\infty$.
Remarkably, the above sign structure of the off-diagonal matrix elements
renders the ground state many-body wavefunction nodeless,
and also leads to the disappearance of the QMC sign problem for the ground
states.

For the finite temperature thermodynamic properties, we use the 
stochastic series expansion (SSE) QMC
method with the  directed loop update algorithm
\cite{sandvik1991,kawashima2004,beard1996,sengupta2002,syljuasen2002}.
This method is usually used for boson systems and 1D fermion systems.
In our case, although the band structure of Eq. \ref{eq:kin} is 
quasi-1D like, the interaction Eq. \ref{eq:int} couples all the chains 
together. 
In particular, the total spin of each chain is not conserved,
and thus its magnetic properties in truly 2D.
Remarkably, we find for this model the sign problem is absent
at finite temperatures in the
entire electron density region $0<n<2$.
In the SSE method, the partition function is expanded as
\bea
Z=\sum_{n=0}^{+\infty}  \frac{\beta^n}{n!} \sum_{\alpha^n_i}
\prod_{i=1}^{n}\bra{\alpha^n_{i}}-H\ket{\alpha^n_{i-1}},
\label{EQ:partition}
\eea
where $H=H_{kin,\pp}+H_{int}$; $|\alpha^n_i\rangle$ runs over the 
set of many-body bases
defined above and $|\alpha^n_n\rangle=|\alpha^n_0\rangle$.
A negative constant is added to the many-body Hamiltonian matrix to
make all of its diagonal matrix elements negative, and then
all the matrix elements of $-H$ become positive.
The grand canonical ensemble is employed to ensure the
ergodicity of the particle number distribution in each chain.
The parameters for the QMC simulations are provided
in Appendix \ref{append:qmc}.

The QMC sign problem does appear in the presence of the $t_\perp$-term
, i.e., Eq. \ref{eq:Hperp},
because electrons become mobile in a two-dimensional manner.
Nevertheless, the QMC simulations can still be performed when the
sign problem is not so severe, which will be presented
in Sect. \ref{sect:tperp}.

\section{Thermodynamic quantities in the off-critical region}
\label{sect:off}
In this section, we present the results of QMC simulations on the 
spin magnetic susceptibility $\chi(T)$ and the compressibility 
$\kappa(T)$ in the off-critical temperature region.
$\chi(T)$ exhibits the celebrated CW law at temperatures well-below 
the kinetic energy scale of the system, while $\kappa(T)$ typically 
weakly depends on temperature.

\subsection{The temperature dependence of spin susceptibility
$\chi(T)$ and compressibility $\kappa(T)$}

\begin{figure}
\includegraphics[height=0.7\columnwidth]
{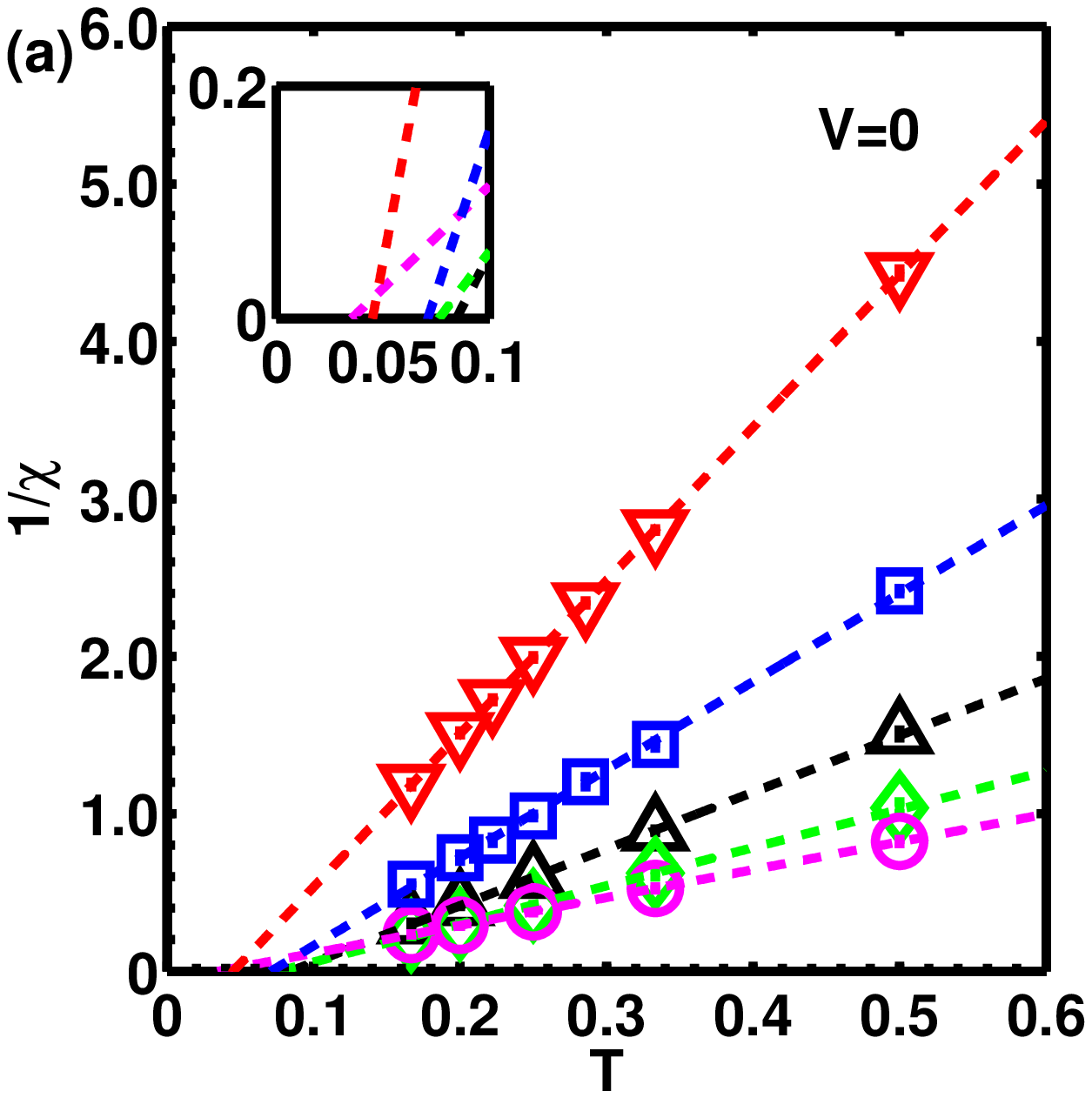}
\includegraphics[height=0.7\columnwidth]
{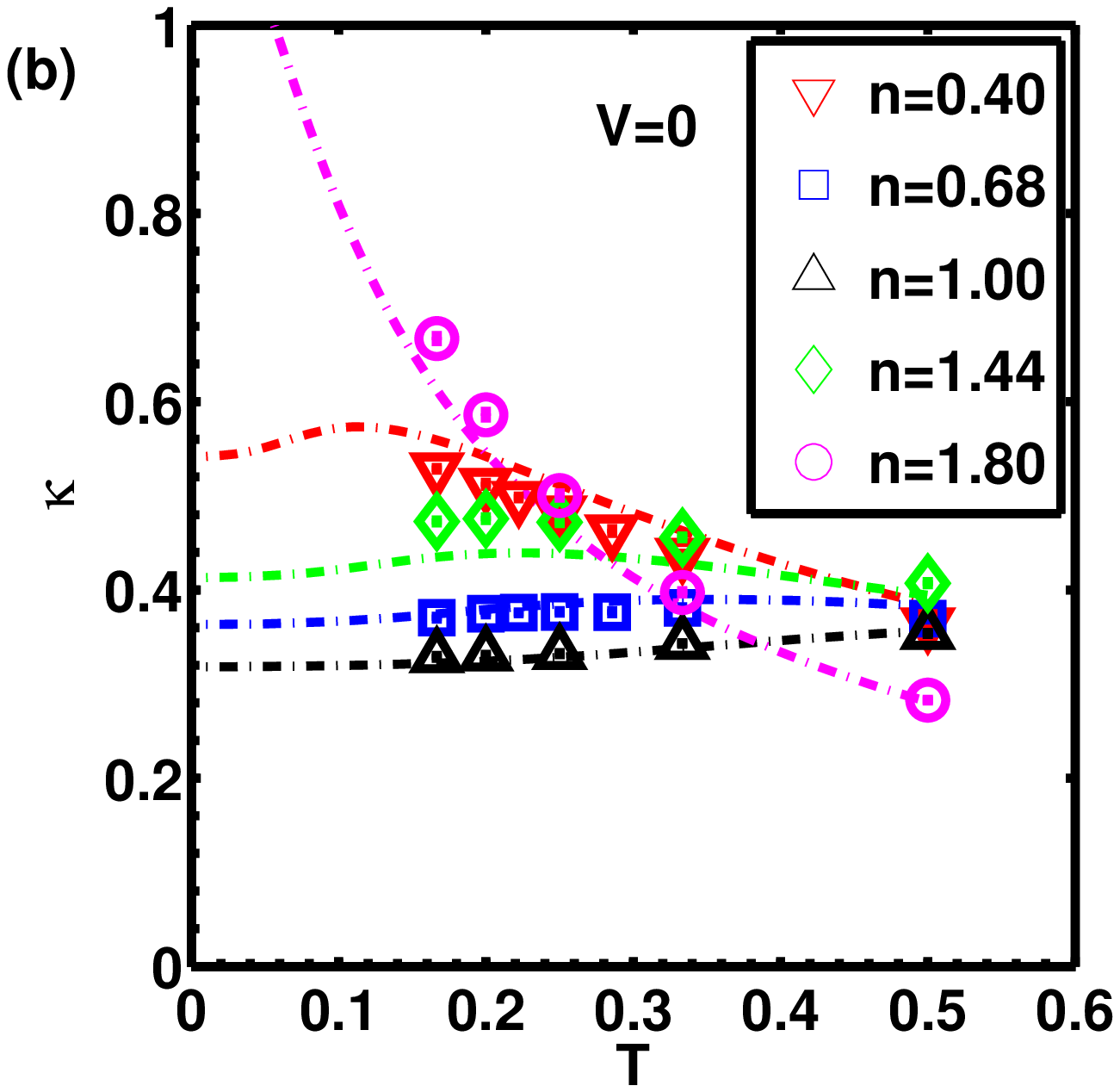}
\caption{($a$) $\chi^{-1}(T)$ exhibits the CW 
law
at different values of $n$.
The inset shows interceptions corresponding to
the mean-field value of Curie temperature $T_{0}$.
($b$) The compressibility $\kappa(T)$ at different values of $n$.
The dashed lines represent $\kappa(T)$ of 1D spinless fermions
at the same densities for comparison.
Values of $n$ in ($a$) and ($b$) are represented by the same legend.
$V=0$ and $J=2$ for both figures.
The error bars of the QMC data are smaller than the symbols.
}
\label{fig:sus}
\end{figure}

The spin susceptibility $\chi$ and compressibility $\kappa$ are two 
fundamental thermodynamic properties in interacting fermion systems in 
the spin and charge channels, respectively.
In usual paramagnetic Fermi liquid states, both $\chi$ and $\kappa$ 
at zero temperature exhibit
the itinerant feature controlled by the density of states at the
Fermi energy.
Furthermore, they are renormalized by interaction effects
characterized by the Landau parameters $F^0_a$ and $F^0_s$ in the spin
and charge channels, respectively.
At finite temperatures much lower 
than the Fermi temperature,
$\chi(T)$ and $\kappa(T)$ are only weakly temperature dependent.
However, in FM metals $\chi(T)$ and $\kappa(T)$ behave dramatically 
differently exhibiting local-moment-like and itinerant features, 
respectively, which will be shown from the QMC simulation results.

Because the total spin is conserved, the spin magnetic susceptibility 
$\chi$ is represented by the equal-time correlation function as
\bea
\chi(T)&=& \lim_{L\rightarrow +\infty}
\frac{\beta}{L^2} \sum\limits_{\vec r_1,\vec r_2}\langle
S_z(\vec{r}_1) S_z(\vec{r}_2)\rangle.
\label{eq:sus_chi}
\eea
The QMC results of $\chi^{-1}(T)$ at $V=0$ are presented in Fig.
\ref{fig:sus} ($a$) in the off-critical region based on the finite size
scalings shown in Appendix \ref{append:thermo}.
For all the values of $n$ presented, $\chi$ exhibits the CW-law
in the off-critical region.
The values of $T_0$ extracted from the linear form $\chi^{-1}(T)$ range
from $0.01$ to $0.1$,
which means that spin remains incoherent at temperatures well below $t_\pp$
(scaled to 1).

It is not surprising that $\chi(T)$ should asymptotically scale as $1/T$
in the high temperature limit  $T\gg T_f$ where $T_f$ is the Fermi 
temperature because in this limit spin channel is completely incoherent. 
Nevertheless, the spin incoherence persists into a much lower 
temperature scale $T_0$ below $T_f$. 
Although $T_0$ is a mean-field energy scale which does not mean the
FM long-range order, it remains important roughly equal to the energy 
cost of flipping an individual electron spin in the ground state.
Due to non-Gaussian fluctuations, the actual FM critical
temperature $T_c$ significantly deviates from $T_0$
defined in Eq. \ref{eq:curie}.
In the current SU(2) invariant case, actually $T_c=0$ 
due to the Mermin-Wagner theorem \cite{mermin1966}.

\begin{figure}
\includegraphics[width=0.7\columnwidth]
{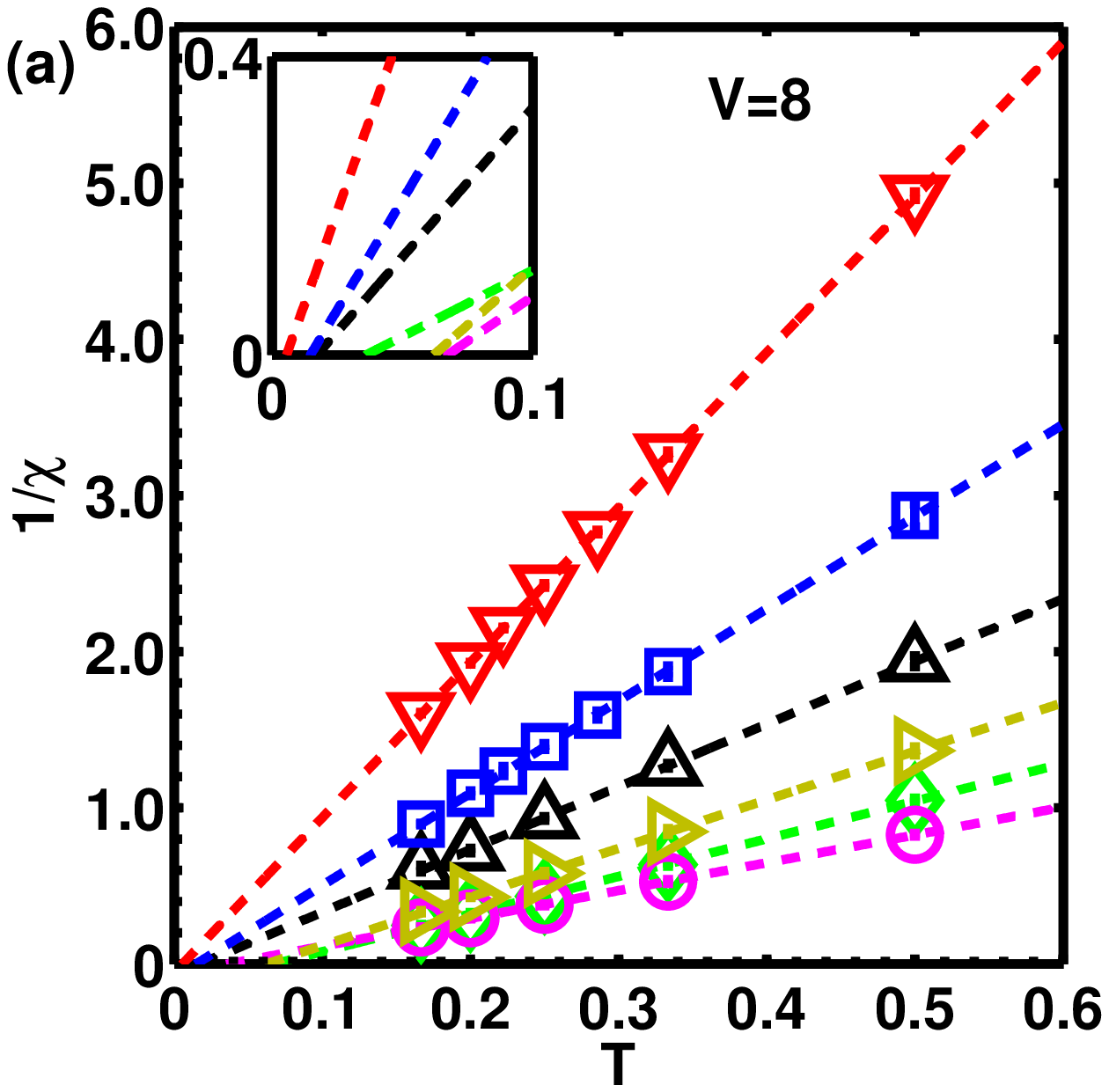}
\includegraphics[width=0.7\columnwidth]
{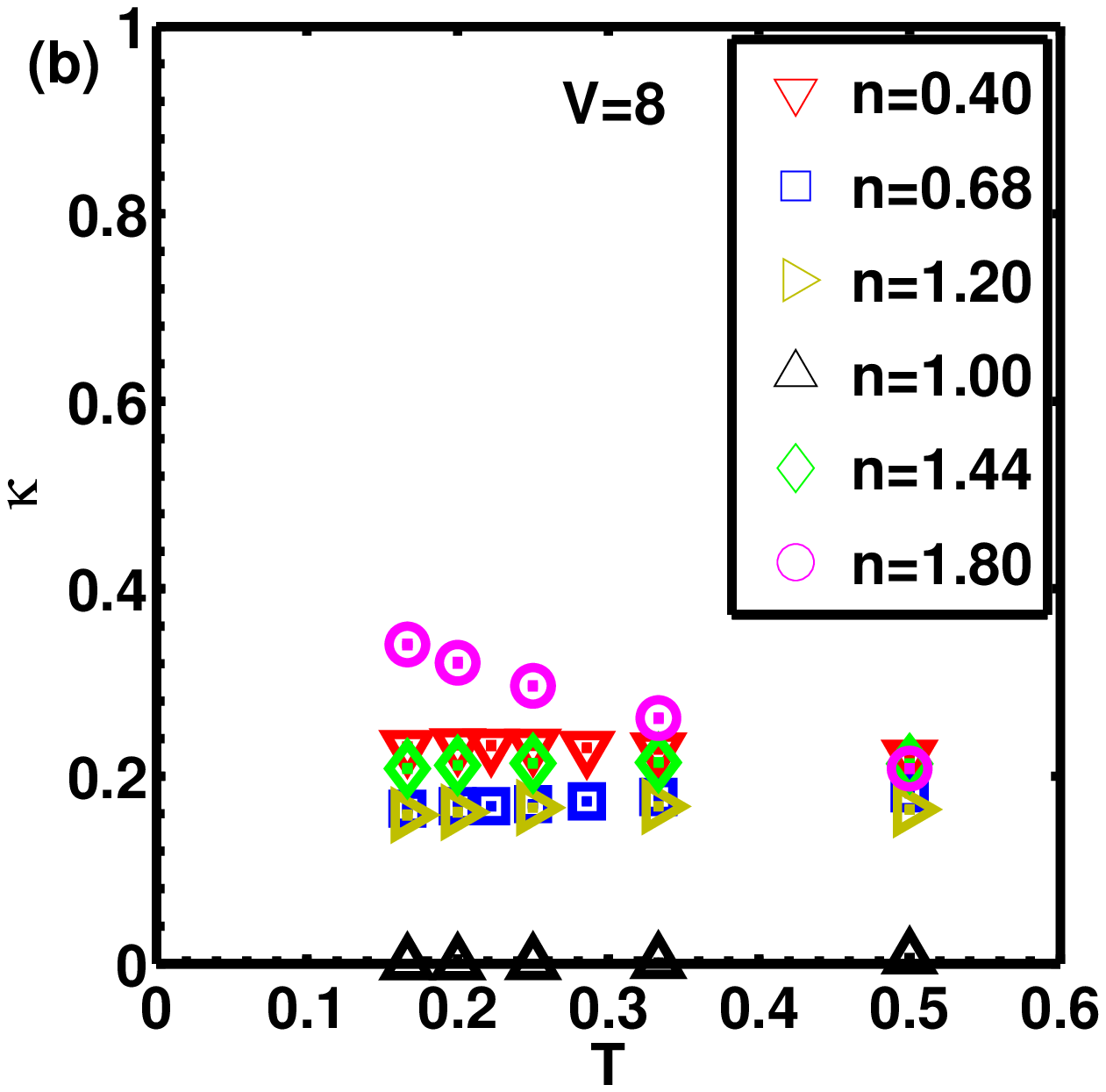}
\caption{($a$) $\chi^{-1}(T)$ and ($b$) $\kappa(T)$ at a large value
of $V=8$ in the temperature regime of $1/6 <T < 1/2$.
Different values of $n$ are shown in the legend in ($b$).
At $n=1$, the system is in the Mott-insulating state, and thus
$\kappa(T)$ drops to nearly zero at low temperatures.
The error bars of the QMC data are smaller than the symbols.
}
\label{fig:susV8}
\end{figure}

The compressibility $\kappa(T)$ reflects the coherence in the
charge channel.
Because the total particle number is a conserved quantity, 
it is also defined as an equal-time correlation function as
 \bea
\kappa(T)&=& \lim_{L\rightarrow +\infty}
\frac{\beta}{L^2} \sum\limits_{\vec r_1, \vec r_2}\langle
n (\vec{r}_1) n (\vec{r}_2)\rangle.
\label{eq:sus_comp}
\eea
The QMC results of $\kappa(T)$ at $V=0$ are presented in 
Fig. \ref{fig:sus} ($b$).
Again $\kappa$ is proportional to $1/T$ in the high temperature incoherent
regime as shown in Eq. \ref{eq:sus_comp},
and it saturates at low temperatures in the metallic phase.
The crossover temperature scale $T_{ch}$ between these two regimes
is typically the chemical potential at zero temperature.
In the usual Fermi liquid state, $\kappa$ is typically the density
of state at the Fermi energy renormalized by Landau parameters.
In our case, the situation is different due to the prominent
FM fluctuations.
At $V=0$, due to the infinite $U$ and the 1D band structure,
$T_{ch}$ is roughly the Fermi temperature of spinless
fermions at the same density.
For most values of $n$ presented in Fig. \ref{fig:sus} ($b$),
$T_{ch}$ is at the order of $t_\pp$, and thus $\kappa$ saturates in
the temperature region presented.
As for the case of a low hole density $n=1.8$, $T_{ch}$ can be estimated
around $0.1$,
and thus $\kappa(T)$ does not saturate yet 
in the simulated temperature region.
Because of the strong FM tendency, the inter-orbital
interaction vanishes at $V=0$ and $\kappa(T)$ can be well fitted
by that of spinless fermions as shown in Fig. \ref{fig:sus} ($b$).

Comparing $\chi(T)$ and $\kappa(T)$,  the spin coherence temperature $T_0$ is 
much lower than the charge coherence temperature $T_{ch}$.
These two distinct coherence temperature scales in spin 
and charge channels are an important feature of FM metals.
A common phenomenological interpretation is to divide electrons 
into two parts: local moments and itinerant electrons which are
responsible for the spin and charge channel behaviors, respectively. 
However, this dividing is artificial for metals when all 
the electrons are itinerant with equivalent band structures such
as in our case. 
Based on the QMC simulations above, we have shown unambiguously
that the CW-law can indeed appear in strongly correlated
systems without local moment formation. 
A similar feature also appears in the CW metal states
\cite{foo2004,merino2006} and the 1D spin incoherent
Luttinger liquids \cite{fiete2007}.
The difference is the behavior of $\chi$ below the spin coherence
temperature $T_0$.
In the case of the CW-metal, $\chi$ saturates exhibiting the
Pauli-like behavior but strongly enhanced by interactions,
and in the 1D case, antiferromagnetic correlations
develops.
In our case, as will be shown in Fig. \ref{fig:critical_Heisenberg}
in Sect. \ref{sect:critical}, $\chi$ evolves into an 
exponential growth as a reminiscence of 
the FM long-range ordered ground state \cite{li2014}. 

Next we consider the effects of a large inter-orbital repulsion $V$
to $\chi^{-1}(T)$ and $\kappa(T)$.
The ground states remain fully spin polarized as proved in
Ref. [\onlinecite{li2014}], and the QMC results of 
$\chi^{-1}(T)$ still exhibit the CW law at all the fillings 
as shown in Fig. \ref{fig:susV8} ($a$).
The most prominent effect of $V$ is the suppression of $\kappa(T)$
at the commensurate filling
of $n=1$ as shown in Fig. \ref{fig:susV8} ($b$), in which the
system is in the Mott-insulating state.
In this case, electrons become local moments due to the 
opening of charge gap.
As a result, $\kappa(T)$ is suppressed to nearly zero at $0<T<0.5$,
which is still small compared to the charge gap at the order of $V$.
In the Mott-insulating ground state at $n=1$, the orbital 
channel can develop the antiferro-orbital ordering with a
staggered occupation of $p_x$ and $p_y$-orbitals. 
The QMC simulation results on the antiferro-orbital ordering
transition are presented in Appendix \ref{append:orbital}.
As $n$ moves away from 1, electrons become itinerant again.
Nevertheless, the values of $\kappa(T)$ at $V=8$ are significantly
suppressed compared to those with the same values of $n$ and $T$ at $V=0$.

\subsection{The density dependences of $T_0(n)$ and the Curie
constant $C(n)$}

\begin{figure}
\includegraphics[height=0.7\columnwidth] {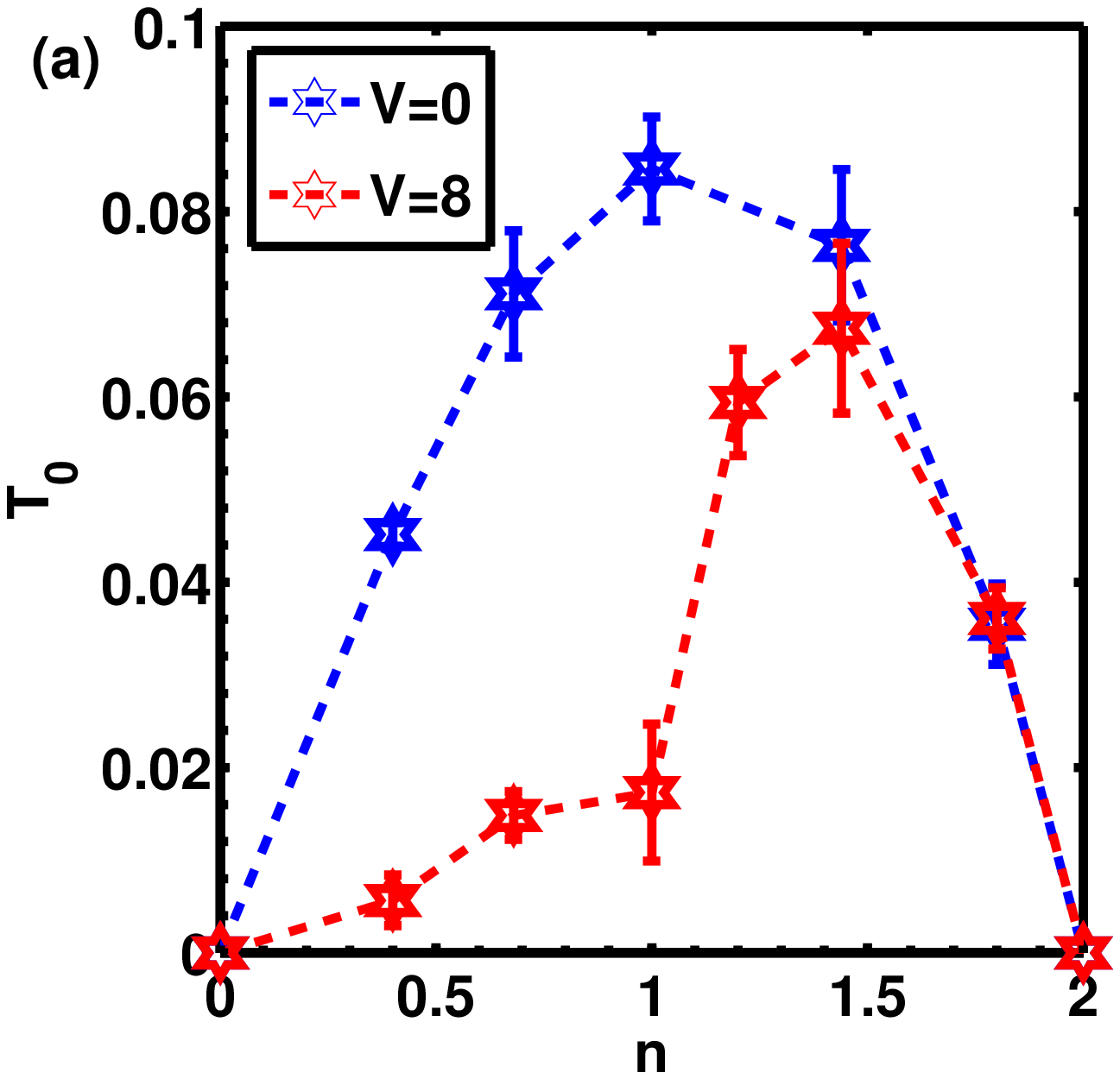}
\includegraphics[height=0.7\columnwidth] {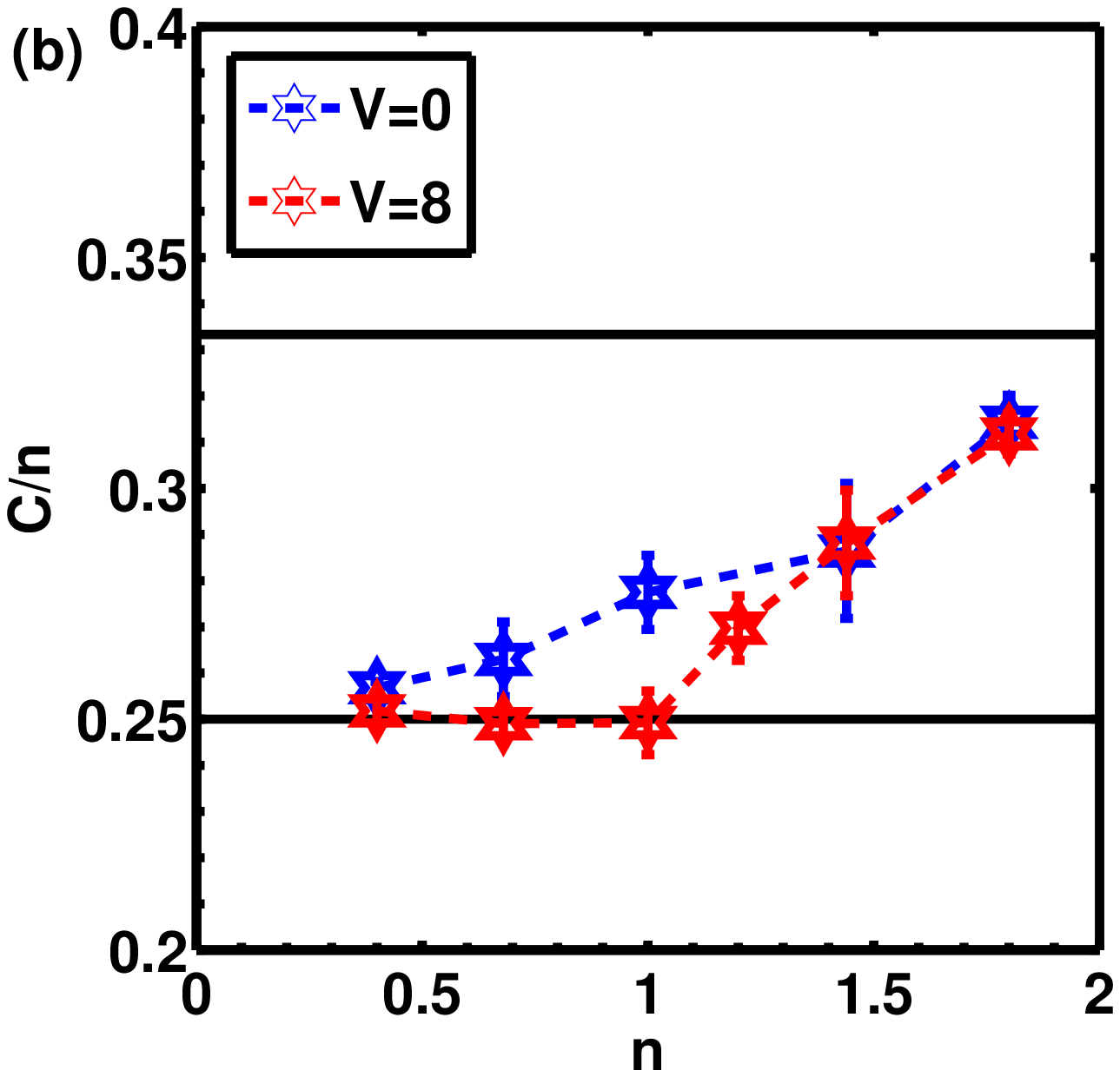}
\caption{($a$) The density-dependence of the Curie
temperature $T_0(n)$ at $V=0$ and $8$ with $J=2$, respectively.
($b$) The density-dependence of the reduced Curie constant:
$C/n$ {\it v.s.} $n$.
The lower and upper bold lines represent the limits of the
spin-$\frac{1}{2}$ and spin-1 moments, respectively.
Plots are based on the results of $\chi(T)$
in Fig. \ref{fig:sus} ($a$) and Fig. \ref{fig:susV8} ($a$).
}
\label{fig:curie}
\end{figure}

The ground state FM survives in all the filling region $0<n<2$,
nevertheless, its robustness against thermal fluctuations varies
at different densities, which reflects through
the density dependences of $T_0(n)$ and $C(n)$.

The relation $T_0(n)$
is presented in Fig. \ref{fig:curie}
($a$) for both cases of $V=0$ and $V=8$.
The FM coherence is built up due to the itinerancy of
fermions \cite{li2014}, thus $T_0$ approaches
zero in both limits of $n \rightarrow 0$ (the particle vacuum)
and $n\rightarrow 2$ (the hole vacuum).
At $V=0$, the maximal $T_0$ appears
around $n=1$ where electrons are most mobile.
$T_0(n)$ at $V=0$ is nearly symmetric with respect
to $n=1$ exhibiting an approximate particle-hole symmetry.
In contrast, it is highly asymmetric at large $V$.
In this case, $T_0$ is strongly suppressed at $0<n<1$, in which both
charge and spin carriers are electrons.
A large $V$ penalizes two electrons occupying the same site,
thus the effectiveness of Hund's rule is suppressed.
After $n$ passes 1, a quick increase of $T_0$ appears because 
extra electrons on top of the Mott  background of $n=1$
can move easily to build up the FM coherence.
$T_0$ reaches the maximum roughly at the middle point between $n=1$ and $2$.
As $n\rightarrow 2$, $T_0$ becomes insensitive to $V$.
In this region, most sites are doubly occupied in the states of 
spin-1 moments, and
holes are itinerant but do not carry spin.
Hole's motion threads spin moments along its trajectory
and aligns their orientations, and this process is not much affected by $V$.
At $V=8 t_\pp$ and $J=2 t_\pp$, the maximal $T_0\approx 0.06 t_\pp$
which appears around $n\approx 1.4$.
In other words, at large values of $V$, there is an approximate 
particle-hole symmetry between $n=1$ to $2$ on the background
of one electron per site.

Next we present the results of the Curie constant $C$.
Assuming the local moment picture, the simple molecule field method
yields $C$ per spin moment as $\frac{1}{3}S(S+1)$
\cite{moriya1985}, where $S$ is the spin magnitude.
In our case mostly itinerant, the magnitudes of $S$ fluctuate:
$C=0$ for the empty site, $\frac{1}{4}$ for the singly occupied site,
and $\frac{2}{3}$ for the doubly occupied site in the spin-1 configuration,
respectively.
We plot the normalized Curie constant $C/n$ {\it v.s.} $n$ in
Fig. \ref{fig:curie} ($b$).
$C/n$ approaches $\frac{1}{4}$ as $n\rightarrow 0$, and $\frac{1}{3}$
as $n\rightarrow 2$ where most sites are spin-1 moments.
Generally, $C/n$ lies between these two limits.
At $V=0$, as $n$ increases, the number of onsite triplets
smoothly increases and so does $C/n$.
Nevertheless, at large $V$, the onsite triplet formation is strongly
suppressed at $0<n<1$, and thus $C/n$ is stuck at $\frac{1}{4}$.
After $n$ passes 1, $C/n$ starts to increase nearly linearly toward 1/3.
As $n\rightarrow 2$, $V$ hardly affects the number of onsite triplets,
and thus $C/n$ also becomes insensitive to $V$
as $T_0$ does.
 
\subsection{The onsite charge fluctuations and spin moments}
\label{sect:onsite}

\begin{figure}
\includegraphics[height=0.7\columnwidth] {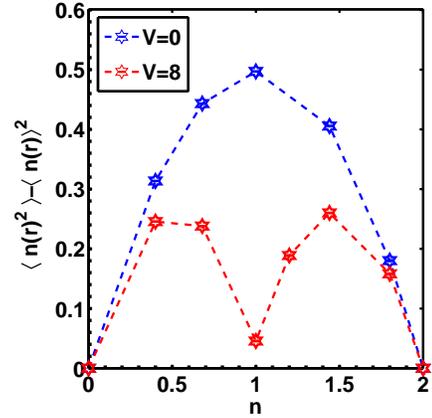}
\caption{The onsite particle number fluctuation
$\delta$ defined in Eq. \ref{eq:ch_fluc}.
The parameters $\beta=6$, $J=2$ and $L=30$.
}
\label{fig:onsite}
\end{figure}

To further clarify the nature of our system whether it is itinerant or 
local-moment-like, we calculate the onsite charge fluctuations and the 
average spin moments below.
The onsite charge fluctuations are defined as
\bea
\delta=\avg{n_i^2}-n^2,
\label{eq:ch_fluc}
\eea
where $n_i$ is the total particle number on site $i$. 
Due to the translation symmetry, $\delta$ is independent of the site
index $i$, and the simulation results are plotted in Fig. \ref{fig:onsite}.
At $V=0$, the charge fluctuations are significant in the entire
filling region except very close to the particle vacuum at $n=0$ and
the hole vacuum at $n=2$.
The maximum is reached at the approximate particle-hole symmetric
point of $n=1$.
The large onsite charge fluctuations clearly reflect the itinerant
nature of the system, which is consistent with the 
compressibility results in Fig. \ref{fig:sus} (b).
When the inter-orbital repulsion $V$ goes large, charge fluctuations
are greatly suppressed near the commensurate filling $n=1$.
In this case, the system becomes local-moment-like, which agrees 
with the vanishing compressibility shown in Fig. \ref{fig:susV8} (b).
Nevertheless, as moving away from $n=1$, the system becomes
itinerant again exhibiting significant onsite charge fluctuations.

We also calculate the square of the $z$-component of the onsite spin moment
$\avg{S_{i,z}^2}$
which equals $\frac{1}{3} \avg{\vec S^2_i}$ since the SU(2) symmetry
is not broken. 
In order to compare with the Curie constant $C/n$, we plot its
values normalized by the filling, i.e. $\avg{S_{i,z}^2}/n$ as presented 
in Fig. \ref{fig:sz2} in Appendix \ref{append:thermo}, which is nearly 
the same as the Curie constant $C/n$ plotted in Fig. \ref{fig:curie} (b).
At $V=0$, $\avg{\vec S^2_i}$ varies smoothly with $n$:
the probable onsite configurations 
include empty, singly occupied (spin-$\frac{1}{2})$, and
doubly occupied (spin-1) states.
At $V=8$ and the commensurate filling $n=1$, $\avg{\vec S^2_i}\approx 
\frac{3}{4}$,
which manifests the formation of the local moment of spin-$\frac{1}{2}$
in consistent with the suppressed charge fluctuations. 
At $0<n<1$,  each site is nearly either empty or singly occupied, 
and thus $\avg{\vec S^2_i} \approx \frac{3}{4} n$.
At $1<n<2$, the probable onsite configurations include the singly
occupied spin-$\frac{1}{2}$ moment and doubly occupied 
spin-1 moment.
Thus the system remains itinerant even in 
at large $V$ when moving away from $n=1$.

\section{The momentum space fermion occupation}
\label{sect:fermi}

\begin{figure}
\includegraphics[height=0.7\columnwidth] {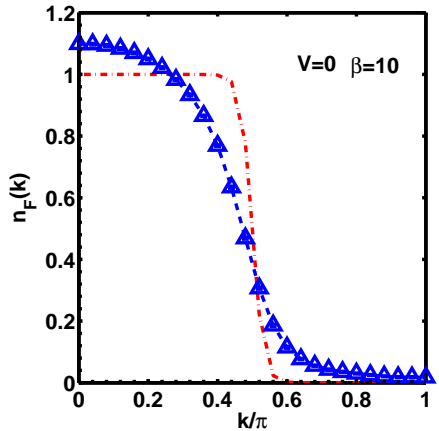}
\caption{
The momentum space distribution $n_F(k)$ ($0\le k\le \pi$)
at $\beta=10$ and $n=1$, and $n_F(k)$ of the non-interacting spinless
fermion (the red dashed line) is plotted for comparison.
The system size $L=50$.
The parameter values are $V=0$ and $J=2$.
The error bars of the QMC data are smaller than the symbol.
}
\label{fig:nf}
\end{figure}

An important feature of the itinerant FM is the fluctuating FM
domains in real space in the paramagnetic phase close to $T_0$.
This prominent FM fluctuations also strongly affect the momentum
space fermion occupation as shown below.
Basically, the fermion occupation functions still resemble
those in the fully polarized systems with thermal broadening.

Because the particle number of each chain is separately conserved,
the momentum space distribution function is essentially 1D-like.
Nevertheless, each chain is not isolated but interacts with
others through multi-orbital interactions, and thus
spin is not conserved separately in each chain.
Without loss of generality,
we define $n_F(k)=\sum_\sigma \avg{p_{x,\sigma}(k) p_{x,\sigma}(k)}$
for a horizontal $x$-chain.
The case of $n=1$ is studied below as a representative, which
is equivalent to $n_x=0.5$ in this $x$-chain.
Its mean-field Curie temperature $T_0 \approx 0.08$ as shown in Fig. \ref{fig:curie} ($a$) before.
The simulated results of $n_F(k)$ are presented in Fig. \ref{fig:nf}
with the periodical boundary condition, and a discussion
on the boundary condition is presented in Appendix \ref{append:thermo}.

We define a reference wavevector as the Fermi wavevector
$k_f^0=\frac{\pi}{2}$ of spinless fermions at the same density.
At a low temperature $T=1/\beta=0.1$ close to $T_0$,
as shown in Fig. \ref{fig:nf}, $n_F(k)$ is only slightly larger 
than 1 even at $k\ll k_f^0$.
It smoothly decays to zero with a half-width approximately equal to $k_f^0$.
$n_F(k)$ is rounded off compared to that of spinless
fermions at the same temperature.
Although $n_F(k)$ does not look much different from that of 
spinless fermions, it is a consequence of strong interactions
because the system is in the paramagnetic state!
The system remains unpolarized with a FM correlation length
$\xi$ at the order of $10 \sim 20$ as estimated in Appendix
\ref{append:thermo}, and the upper bound of $n_F(k)=2$
as $k\rightarrow 0$.

The above result implies that the phase space for thermal 
fluctuations is not restricted to a small region close to 
$\pm k_f^0$, and thus its entropy capacity is enhanced.
It is consistent with the real space picture of
fluctuating FM domains as $T$ approaches $T_0$.
This is highly non-perturbative showing the
power of the QMC simulations.

\section{The low temperature critical region}
\label{sect:critical}
\begin{figure}
\includegraphics[height=0.7\columnwidth]{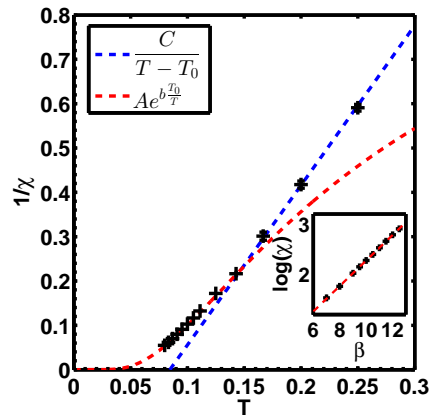}
\caption{The SU(2) invariant model in the critical region:
$\chi^{-1}(T)$ at $n=1$, $J=2$ and $V=0$.
It crosses over from the CW 
law to the exponential form of
$\chi(T)= A e^{b\frac{T_0}{T}}$.
The inset shows the linear scaling of $\ln \chi(T)$ v.s. $\beta$.
}
\label{fig:critical_Heisenberg}
\end{figure}

\begin{figure}
\includegraphics[width=0.75\linewidth] {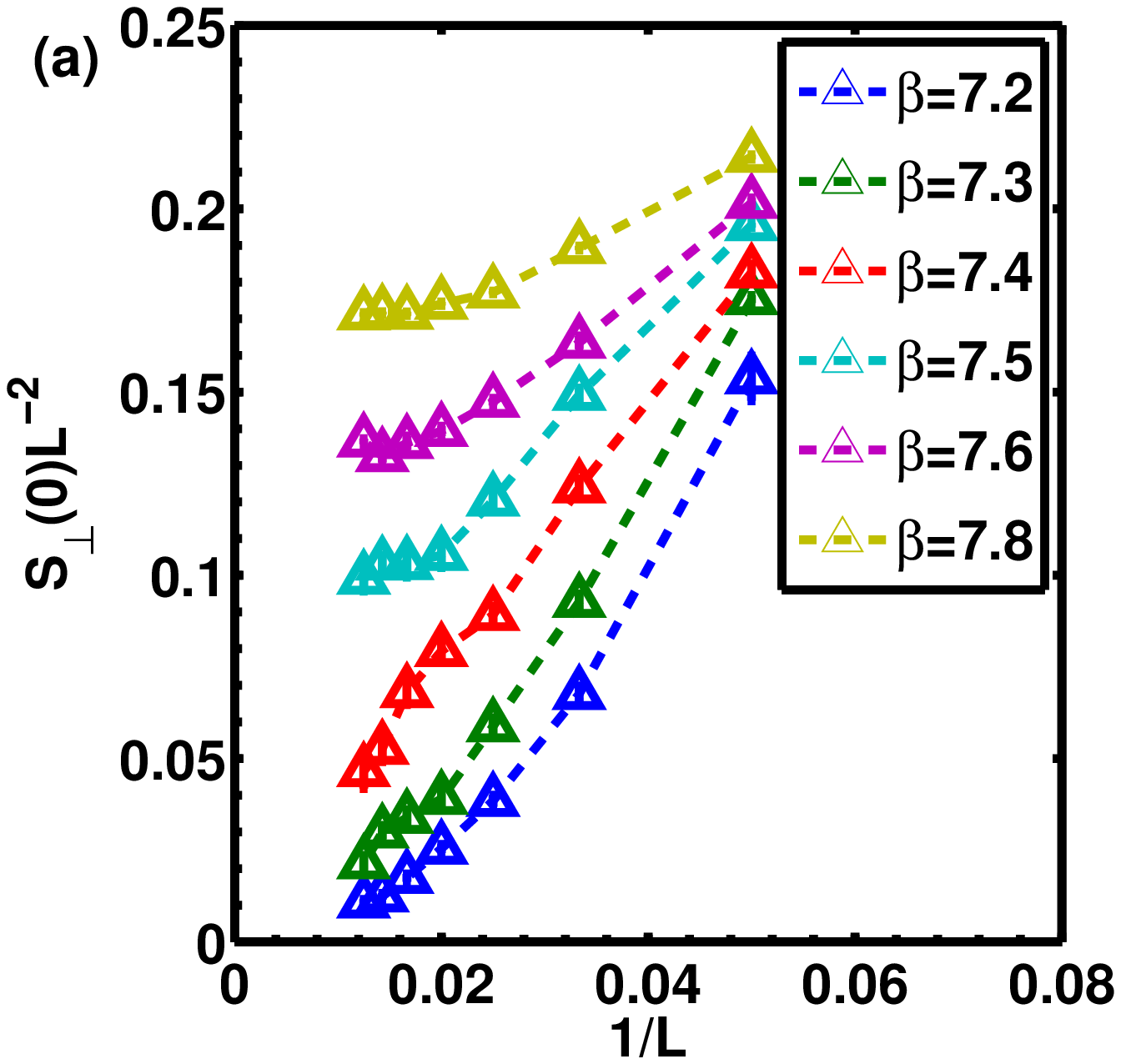}
\includegraphics[width=0.75\linewidth] {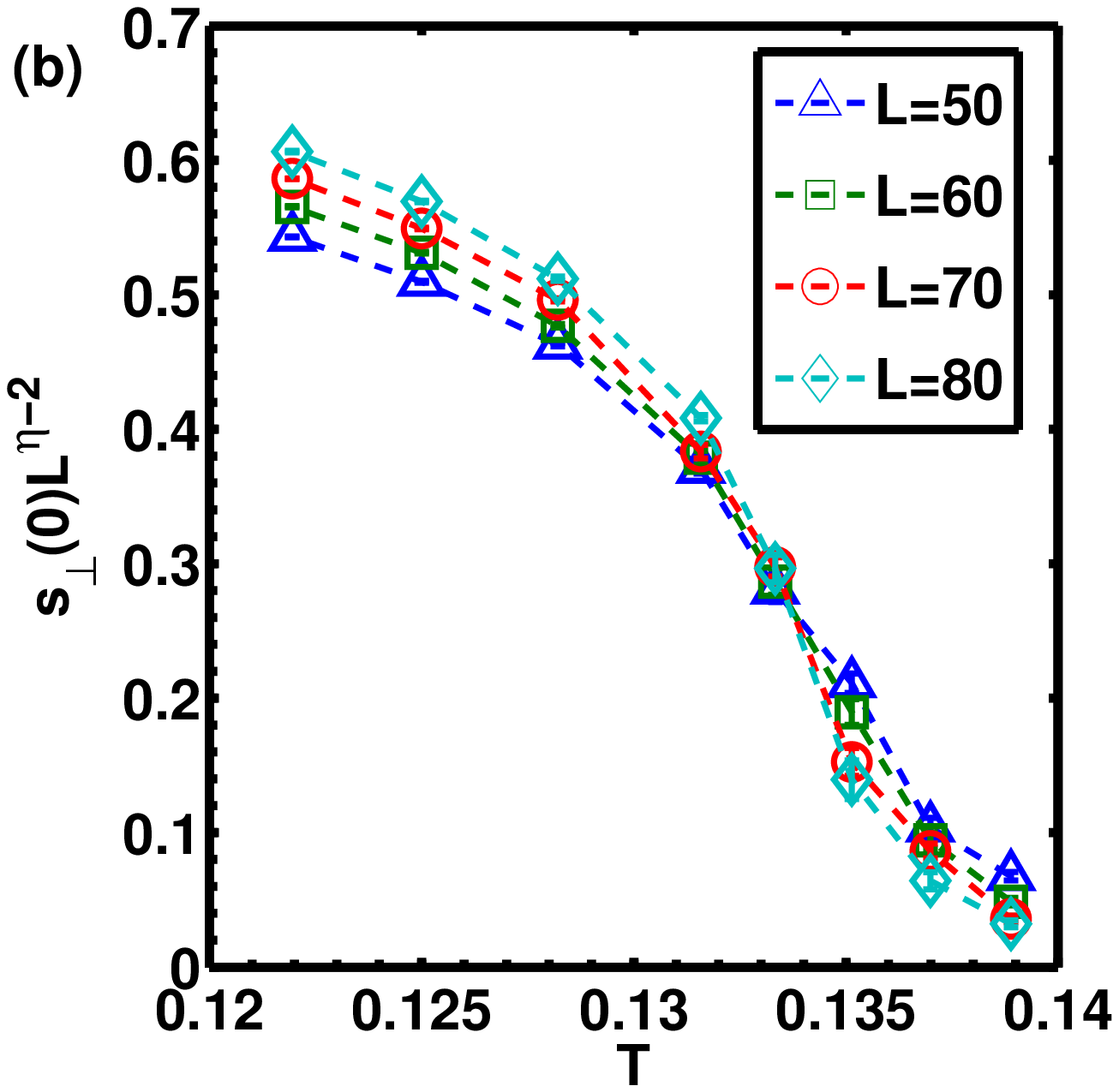}
\includegraphics[width=0.75\linewidth] {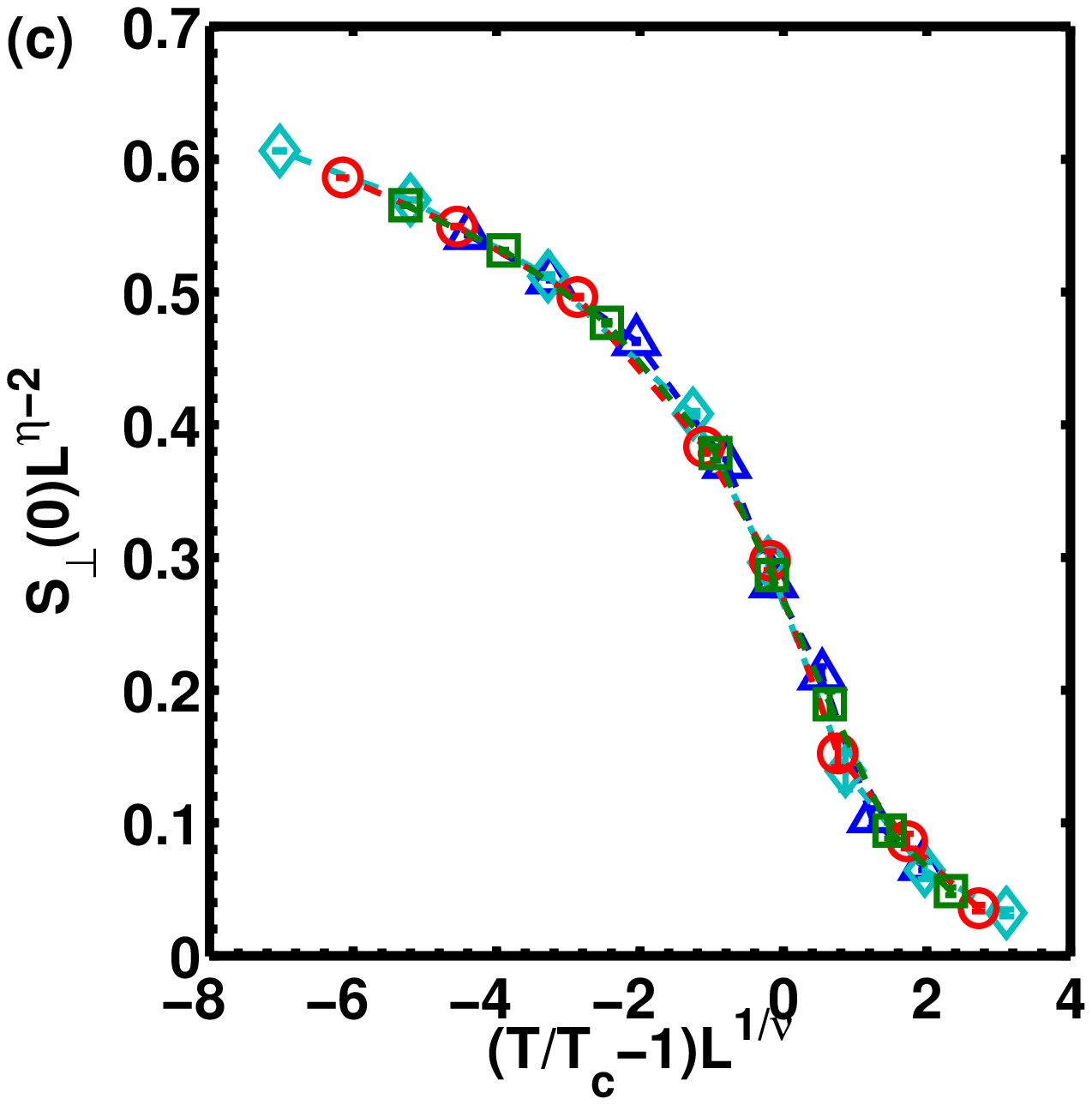}
\caption{The FM long-range ordering of the Ising symmetric model
with parameters $J_\perp=2J_\pp=4$, $n=1$, and $V=0$.
($a$) The finite size scaling of $S_\perp/L^2$.
The FM critical temperature is extracted as $T_c=1/\beta_c=0.134\pm 0.002$
with $\beta_c\approx 7.4\sim 7.5$.
($b$) The critical scaling: $S_\perp L^{-2+\eta}$ v.s. $T$ with
$\eta=\frac{1}{4}$.
The crossing of curves at different values of $L$ yields
$T_c\approx 0.134$.
($c$) The data collapse of the scaling form $S_\perp(0)L^{-2+\eta}=
f((T-T_c)L^{1/\nu})$ fitted by the parameters of
$\nu=1$, $\eta=\frac{1}{4}$ and $T_c=0.1337$.
}
\label{fig:critical_ising}
\end{figure}

So far we have discussed the FM properties in the off-critical region.
In this section we will further study the magnetic critical 
behavior through QMC.
The FM order parameter is a conserved quantity, and thus
there are no quantum fluctuations, however, in 2D thermal 
fluctuations are so strong that long-range FM ordering cannot 
appear at any finite temperatures for SU(2) symmetric 
models \cite{mermin1966,zaleski2008}.
Nevertheless, magnetic properties still behave qualitatively
differently in the off-critical and critical regions.
We will also consider the model in the Ising class
in which true FM long range ordering can appear, and 
determine the renormalized Curie temperature $T_c$.

In spite of the quasi-1D band structure, the magnetic properties
of our model are intrinsically 2D because Hund's interaction
couples spins of different chains together and the total spin
of each chain is not  separately conserved.
In Fig. \ref{fig:critical_Heisenberg}, we present the crossover
of $\chi^{-1} (T)$ from the off-critical region to the critical
region based on the finite size scaling presented in Appendix
\ref{append:thermo}.
Although there is no distinct phase transition between the
off-critical and critical regions, the temperature dependence
of $\chi(T)$ changes qualitatively.
The clear deviation from the CW 
law starts from
$T\sim T_0=0.08$.
In the critical region, the FM order parameter already develops
a non-zero magnitude, and its directional fluctuations are described
by the O(3) non-linear $\sigma$-model.
The FM correlation length increases exponentially as approaching
zero temperature.
$\chi(T)$ evolves to the exponential form fitted by
$\chi = A e^{b\frac{T_0}{T}}$ \cite{arovas1988,takahashi1987}, and
the result in Fig. \ref{fig:critical_Heisenberg} shows $b=3.1\pm 0.3$
at $n=1$, $V=0$, and $J=2$.

In order to obtain the FM long-range order, we modify Hund's 
coupling of Eq. \ref{eq:int} to reduce its symmetry from the SU(2) to
the Ising class: We introduce $J_\pp$ and $J_\perp$ for the spin components
in the $xy$-plane and along the $z$-direction, respectively,
and choose $J_\perp>J_\pp$.
The $z$-component FM structure factor is defined as
$S_\perp(T,L)=T\chi(T,L)$.
For the case presented in Fig. \ref{fig:critical_ising} ($a$),
the finite size scaling of $S_\perp(T,L)/L^2$ yields the critical
temperature $T_c\approx 0.134$.
This result is also checked from the scaling in the critical region
in Fig. \ref{fig:critical_ising} ($b$) and ($c$).
$S_\perp L^{-2+\eta}$ {\it v.s.} $T$ is plotted with $\eta=\frac{1}{4}$
from the anomalous dimension of the 2D Ising universal class.
The crossings of curves yield the value of $T_c$ consistent with
that of the previous scaling.
Furthermore, a good data collapse is achieved by employing
the scaling form
\bea
S_\perp L^{-2+\eta}= f((T-T_c) L^\frac{1}{\nu})
\eea
with $\nu=1$ of the 2D Ising class.

In Appendix \ref{append:ising}, the mean-field value
$T_0\approx 0.20$ is extracted based on the extrapolation
of the CW behavior.
Compared to the mean-field value $T_0$, $T_c$ is about $67\%$
of $T_0$ as a result of the critical non-Gaussian fluctuations.
For the 2D Ising mode with only nearest neighbor coupling
 on the square lattice, the Onsager solution gives
rise to $T_c=2/\ln (\sqrt 2+1)\approx 2.269$ which is $57\%$
of the Bragg-Williams mean-field results $T_0=4$.
Thus the critical fluctuation strength of the case presented
in Fig. \ref{fig:critical_ising} is weaker compared to
that in the 2D Ising model in spite of the effect of the
transverse component $J_\pp$.
This is due to the itinerant nature of our model such that
the effective FM coupling is beyond two nearest neighboring sites.


\section{Discussions on the $t_\perp$-term and the finite $U$}
\label{sect:discussion}

\begin{figure}
\includegraphics[height=0.6\columnwidth]{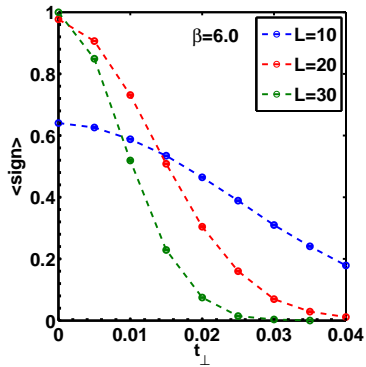}
\caption{The average of the fermion sign v.s. $t_\perp$ 
at different values of $L$.
The parameters are $V=0$, $J=2$, $n=1$, and $\beta=6$.
}
\label{fig:sign}
\end{figure}

\begin{figure}
\includegraphics[height=0.6\columnwidth]{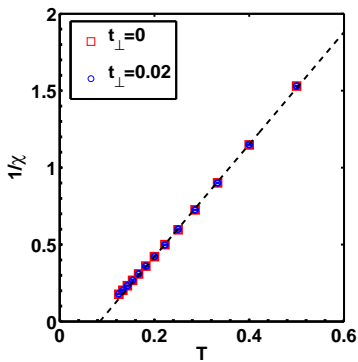}
\caption{$\chi^{-1}(T)$ at $t_\perp=0$ and $0.02$, respectively,
with the system size $L=20$.
Periodical boundary conditions are used.
The parameters are $V=0$, $J=2$, $n=1$.
}
\label{fig:sign_sus}
\end{figure}

In this section, we discuss the situations when the conditions for the 
absence of sign problem are loosed, including the presence of a small 
transverse hopping $t_\perp$-term as shown in Eq. \ref{eq:Hperp}, and 
the case of finite values of $U$.

\subsection{QMC simulations with small transverse hopping term}
\label{sect:tperp}


The presence of the $t_\perp$-term enables electrons moving in the entire 
two-dimensional lattice, thus the fermion sign problem does appear.
Nevertheless, the sign problem is not severe at small values of
$t_\perp$, such that QMC simulations can still be performed.
In Fig. \ref{fig:sign}, the average of sign is calculated from $t_\perp=0$ 
to 0.05 at small and intermediate sample sizes with $\beta=6$.
We use the periodical boundary condition for the entire system, which is 
different from the boundary condition used in previous calculations 
in order to eliminate the sign problem when electrons
hop across the boundary.
The previous boundary condition is feasible at $t_\perp=0$  because particle
number in each chain is conserved as explained in Sect. \ref{sect:sign}.
Now under the periodical boundary condition, even at $t_\perp=0$
the sign is not positive definite: when one electron hops across the 
boundary, if the fermion number in that chain is an even number, 
the matrix element acquires an extra sign.
This boundary effect is more prominent at small sample sizes 
(e.g. $L=10$) but already becomes negligible at intermediate 
sample sizes, say, $L>20$.
As $t_\perp$ deviates from 0, the 2D motion of electrons suppresses 
the average sign and it drops more rapidly at larger sample sizes.

We have simulated the spin susceptibility and presented its inverse 
$\chi^{-1}(T)$ in Fig. \ref{fig:sign_sus} with $t_\perp=0.02$.
The results at $t_\perp=0$ under the periodical boundary condition 
are also plotted for a comparison.
An intermediate sample size ($L=20$) is used and the simulation 
is performed from the high to intermediate temperature regions.
The results at $t_\perp=0.02$ are nearly the same as those at 
$t_\perp=0$, which still exhibit the CW behavior. 
At the lowest temperature simulated $\beta=6$, the average 
sign at $t_\perp=0.02$ is already significantly below 1.
Nevertheless, the difference between $\chi^{-1}(T)$'s at $t_\perp=0$
and $0.02$ remains negligible. 
These results show that the magnetic properties 
are not so sensitive to $t_\perp$ when $t_\perp/t_\pp\ll 1$.

Certainly, when $t_\perp$ reaches the same order as $t_\pp$,
the band structure will become genuinely two-dimensional.
In this case, the previous ground FM theorem does not apply, and 
a quantum phase transition is likely to occur
from the FM to paramagnetic ground states. 
Unfortunately, the sign problem will be very severe and thus
reliable QMC simulations cannot be performed.
It would be interesting to further develop other analytic and
numeric methods to investigate this problem.

\subsection{The effect of the finite $U$}

As explained in Sect. \ref{sect:sign}, the many-body bases for
simulations, which are also used for the proof of 
FM ground state theorems in Ref. \cite{li2014}, are 
constructed by ordering electrons according 
to their locations along one chain by another regardless of their 
spin configurations. 
This set of bases are convenient to accommodate to the spin-flip
term of Hund's coupling to be free of the sign problem,
nevertheless, finite $U$ does cause this problem. 
If $U$ is finite, states with doubly occupied orbitals are allowed, 
and electrons with opposite spins can exchange their locations
which causes the sign problem.
We will defer the QMC simulations for this case to a later
publication, but briefly analyze the physical effect below. 

Basically, a large but finite $U$ introduces an antiferromagnetic (AFM) 
energy scale of $J_{AF}=4t^2_\pp/U$ for two electrons 
lying in adjacent sites in the same chain. Its effect in the low electron 
density region is unimportant but becomes important in the limit 
of $n\rightarrow 2$ 
in which most sites are occupied as spin-1 moments. In this region, 
the FM energy scale $T_0(n)$ is suppressed because of the low density of 
mobile holes and finally it becomes weaker than $J_{AF}$.
Consequently, we expect a ground state phase transition at a critical 
density $n_c$ close to $n=2$, which marks a transition from the FM ordering
at $n<n_c$ to the AFM ordering at $n_c<n<2$.


\section{Experiment realizations}
\label{sect:exp}

The QMC simulations presented above are not only of academic interests 
but also provide  new directions to explore new FM materials in various 
physical systems, including both the ultra-cold atom optical
lattices and the strongly correlated transition metal oxides.

Recently, the study of itinerant FM states has become a research focus 
in ultra-cold cold atom physics \cite{duine2005,Jo2009,zhangSZ2010,
berdnikov2009,pekker2011a,chang2010,cui2014,pilati2014}.
However, so far it is still in debate whether the experiment results
based on the upper branches of the Feshbach resonances have shown
the existence of itinerant FM or not.
Our work suggests a new direction for the further experiment exploration 
of itinerant FM in the high orbital bands in optical lattices.
Our band Hamiltonian can be accurately implemented in the $p$-orbital 
band in the ultra-cold atom optical lattices 
\cite{isacsson2005,liu2006,wang2008}.
Due to the anisotropy of $p$-orbital orientation, the transverse
$\pi$-bonding amplitude $t_\perp$ is usually much smaller than 
the longitudinal $\sigma$-bonding $t_\pp$. 
The ratio of $t_\perp/t_\pp$ decreases as increasing the optical 
potential depth $V_0$.
As shown in Ref. \cite{isacsson2005}, as $V_0/E_R=15$ where $E_R$ is the
recoil energy of the laser forming the optical lattice,
$t_\perp/t_\pp\approx 5\%$, such that we can neglect the $t_\perp$ term
in Eq. \ref{eq:kin}.
Furthermore, the interaction strength is also tunable in optical lattices 
by simply varying laser intensities, and the strong coupling regime 
can be reached. 
A variation study based on the Gutzwiller projection also shows that
the ground state FM may start from intermediate coupling
strength \cite{wang2008}.
Our simulations on the thermodynamic properties provide important
guidance for future experiments.

Our work is also helpful for the current effort of searching for 
novel FM materials in transition metal oxides, in particular, 
in systems with the $t_{2g}$-orbital bands, i.e., $d_{xz}, d_{yz}$, 
and $d_{xy}$ bands, with the quasi-2D layered structure.
In fact,  FM has been observed experimentally
in the ($001$) interface of $3d$-orbital transition-metal 
oxides such as SrTiO$_3$/LaAlO$_3$ 
\cite{lilu2011,bert2011,michaeli2012,chen2013,banerjee2013}, which
has been a recent research focus in condensed matter physics.
The dispersions of $d_{xz}$ and $d_{yz}$-orbital bands are also highly 
anisotropic,{\it i.e.}, the longitudinal bonding parameter $t_\pp$
is much larger than the transverse one $t_\perp$, as described in
Eq. \ref{eq:kin}  by replacing $p_{x(y)}$ with $d_{x(y)z}$.
The onsite repulsive interaction of the $3d$-electrons are particularly 
strong, such that the projection of doubly occupied orbitals is a
good approximation. 

Even though there is an additional quasi 2D-$d_{xy}$-orbital band in the 
SrTiO$_3$/LaAlO$_3$ interfaces, which is presumably paramagnetic 
by itself, it is conceivable that the overall system remains FM as shown in 
experiments and our results still apply qualitatively.
The reason is that the quasi-1D bands $d_{x(y)z}$ do not hybridize
with the quasi-2D $d_{xy}$ band by the nearest neighboring hopping
due to  their different parity eigenvalues under the reflections with 
respect to $xy$, $yz$ and $zx$-planes, respectively.
It is a good approximation that the particle numbers in the $d_{xy}$-band 
and in the $d_{x(y)z}$ bands are separately conserved, and they 
only couple through interactions.
The coupling is ferromagnetic by nature due to Hund's rule. 
Since the quasi-1D bands by themselves are already FM in the 
strong coupling regime, their coupling
to the paramagnetic $d_{xy}$-band is like to use a permanent ferromagnet
to polarize a paramagnet, and it is conceivable that 
in overall the ferromagnetism is enhanced.


\section{Conclusions}
\label{sect:conc}


In summary, we have non-perturbatively investigated the thermodynamic 
properties of an unambiguous itinerant FM system with multi-orbital 
structures through the method of the SSE QMC.
The simulations are proved to be sign problem free in all the 
electron density region, and thus reliable numerical results 
can be obtained at high numeric accuracy.
Due to the nature of asymptotic exactness of our simulations,
they provide a solid reference point for the study of the strong
correlation effects of the thermodynamic properties of itinerant 
FM systems.
There is a wide temperature region $T_0<T<T_{ch}$, in which
the spin channel is incoherent without
local moments existing as a priori, while the charge channel
exhibits the metallic behavior.
The spin magnetic susceptibility exhibits the CW law in the
off-critical region as a result of strong correlations.
It further crosses over to the exponential
growth in the critical region.
The compressibility is weakly temperature dependent
and saturates to its zero temperature value.
The true FM long-range transition appears when the symmetry
class is reduced from SU(2) to Ising.
The finite size scaling in the critical region gives rise to
an accurate determination of the FM transition temperature.
Our work is also closely related to the experiment efforts of
searching for novel FM states of matter in both ultra-cold
atom optical lattices and in the $3d$ transition metal
oxide materials.

\acknowledgements
C. W. is grateful for
the hospitality of Center of Mathematical Sciences and
Applications at Harvard University where part of the work was
done.
S. X. and C. W. are supported by the NSF DMR-1410375 and AFOSR
FA9550-14-1-0168.
Y. L. is grateful for the support from the Princeton Center for
Theoretical Science.
S. X., Y. L. and C. W. thank D. Arovas, X. Dai,
J. E. Hirsch, M. Randeria, A. W. Sandvik, H. Shao,
L. J. Sham, D. J. Singh, N. Trivedi, and Lu Yu 
for helpful discussions.
All the simulation was performed on Tiger Cluster in Princeton.
C. W. acknowledges the support from the
National Natural Science Foundation of China (11328403), and
the support from the Presidents Research Catalyst Awards of 
University of California.
Y. L. thanks the hospitality of the
Aspen Center of physics under the
support of the NSF Grant No. PHYS-1066293.

\appendix

\section{Parameters for QMC simulations}
\label{append:qmc}

\begin{figure}[h]
\centerline{
\includegraphics[height=0.5\columnwidth,width=0.5\columnwidth] {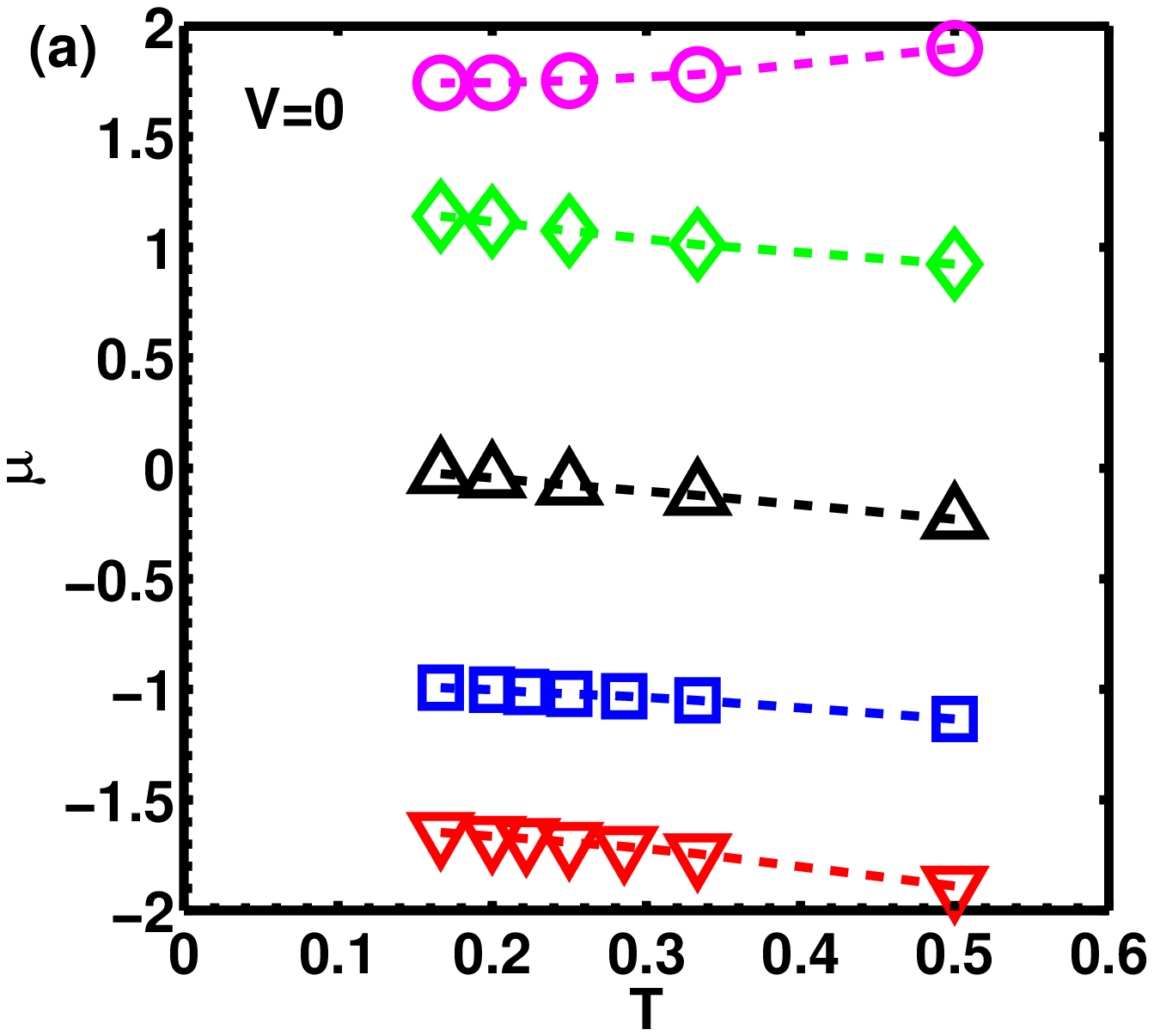}
\includegraphics[height=0.5\columnwidth,width=0.5\columnwidth] {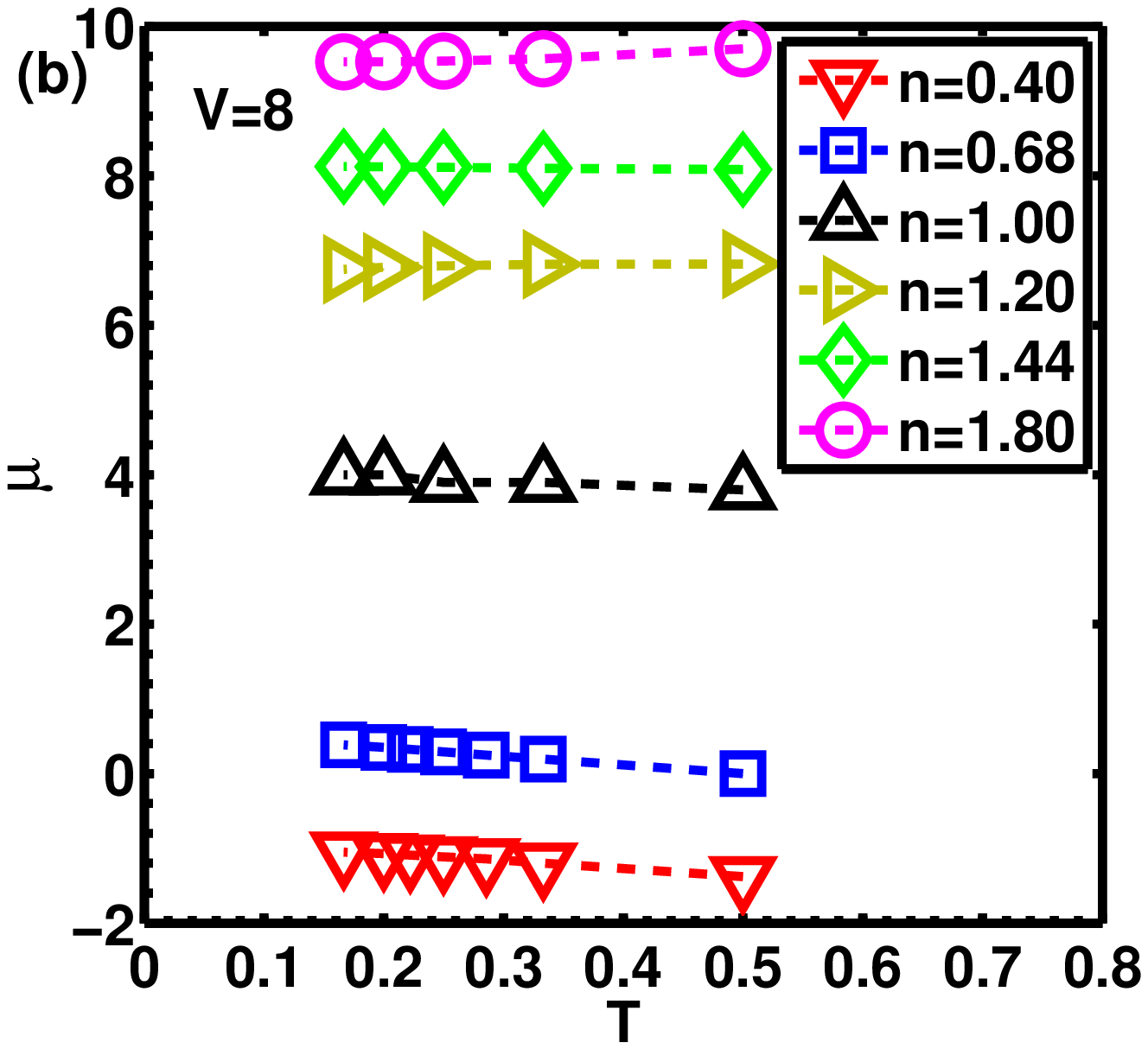}
}
\caption{The chemical potential $\mu(T)$ is tuned to maintain
the filling $n$ unchanged at all temperatures.
The system size is $L=8$. }
\label{fig:filling}
\end{figure}

\begin{figure}
\centerline{\includegraphics[height=4.2in,width=\columnwidth]
{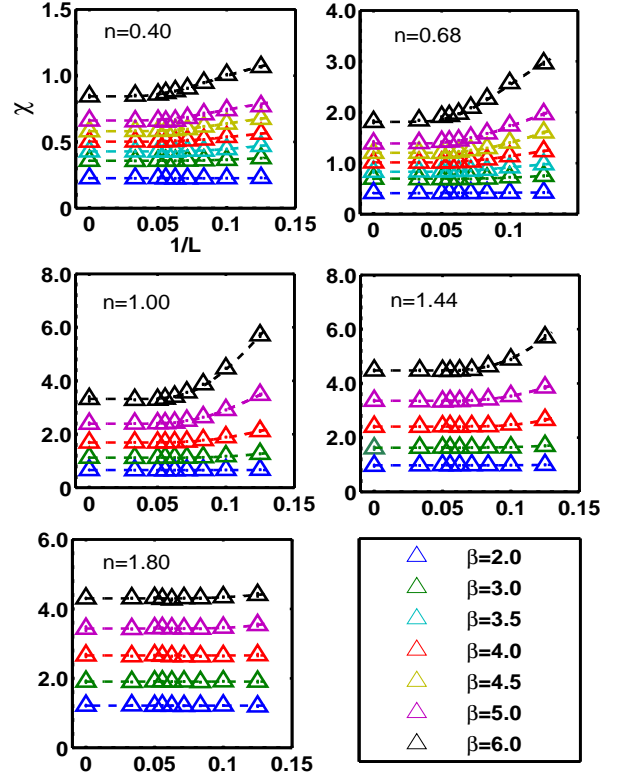}}
\caption{The finite size scalings for $\chi(T,L)$ v.s. $1/L$
with the fitting $\chi(T,L)=aLe^{-\frac{L}{b}}+\chi(T)$.
Values of $\beta=1/T$ are presented in the legend.
The parameter value at $V=0$ and $J=2$.
}
\label{Fig:scaling_v=0}
\end{figure}

We have used SSE method of QMC to simulate the Hamiltonian
Eq. \ref{eq:kin} plus Eq. \ref{eq:int} in the main text with
the directed loop algorithm
\cite{sandvik1991,kawashima2004,beard1996,sengupta2002,syljuasen2002}.
A Monte Carlo step is defined as a diagonal vertex update followed by
many directed off-diagonal loop updates to ensure that most of vertex
legs are visited by the worm-head.
The simulated system size of the square lattice is $L\times L$ with
the values of $L$ given in each figure.
The simulations are run in parallel on 16 cores.
On each core $1.5\times10^5$ warm-up steps are used, and for a typical
data point we use $10^5$ QMC steps and perform $10^4$ measurements.
For the simulation with the largest system size ($L=80$),
$3.0\times10^5$ warm-up steps and $2.0\times10^5$ QMC steps are used.

In order to maintain ergodicity, the directed loop algorithm is
carried out in the grand canonical ensemble in which the chemical
potential $\mu$ is the characteristic variable.
Nevertheless, in realistic systems, it is more natural to fix
the average fermion number per site $n$ rather than to fix $\mu$.
For example, in a system with a fixed average value of $n$, $\mu(T)$
changes as varying the temperature $T$.
Therefore, in presenting the simulation results for a fixed value of
$n$, we have carefully adjusted $\mu$ to maintain $n$ invariant.
The obtained relations of $\mu(T)$ at various values of $n$ with the
system size $L=8$ are plotted in Fig.
\ref{fig:filling}.
The dependence of $\mu$ on $L$ is weak, and thus we use
the same set values of $\mu(T)$ for even larger
sample sizes except the case of $V=0$ and $n=0.40$.
In this case, the finite size effect of $\mu$ is relatively
strong, and we calculate $\mu(T)$ at $L=10$ and use it
for larger sample sizes.
By this method, we can maintain the values of $n$ invariant
within the error of $\Delta n=0.006$ for all the simulations.

\begin{figure}[h]
\centerline{
\includegraphics[height=4.2in,width=\columnwidth] {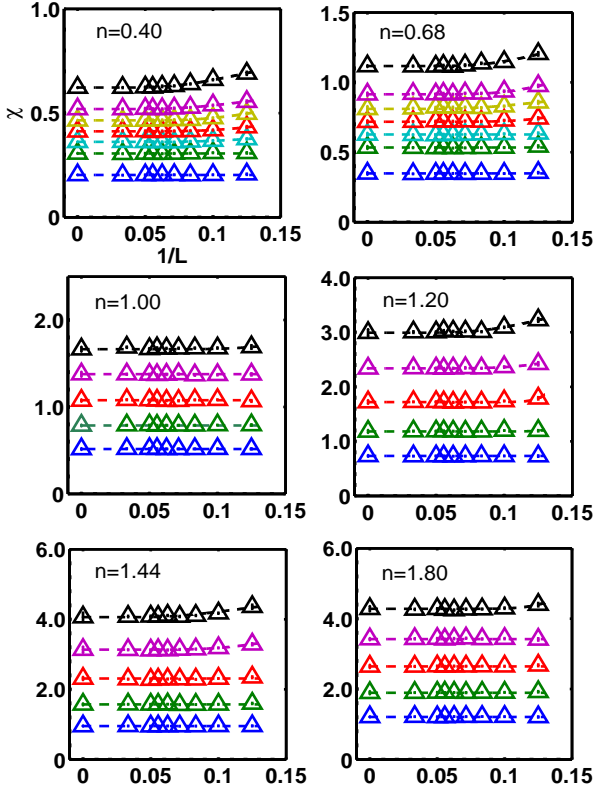}
}
\caption{The finite size scaling of $\chi(T, L)$ v.s. $1/L$ for $V=8$.
Scaling ansatz
$\chi(T,L)=aLe^{-\frac{L}{b}}+\chi(T)$ is used.
The legend is the same as that in Fig. \ref{Fig:scaling_v=0}.
}
\label{Fig:scaling_v=8}
\end{figure}

\begin{figure}
\centerline{
\includegraphics[height=0.7\columnwidth,width=0.6\columnwidth]
{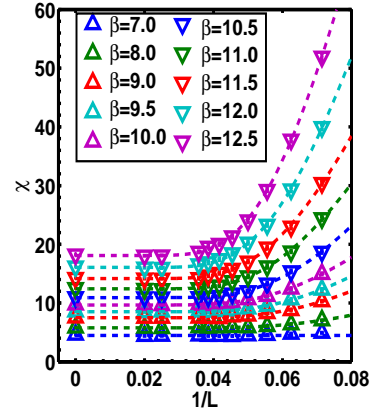}}
\caption{The finite size scaling of $\chi(T,L)$ at $n=1$, $V=0$,
and $J=2$ in the low temperature critical region with
the scaling ansatz $\chi(T,L)=aLe^{-\frac{L}{b}}+\chi(T)$ used.
Values of $\beta=1/T$ are presented in the legend.
}
\label{fig:hei_critical}
\end{figure}

\section{More information on thermodynamic properties}
\label{append:thermo}

In Fig. \ref{Fig:scaling_v=0} and Fig. \ref{Fig:scaling_v=8}, the
finite size scaling of $\chi(T, L)$ are plotted with $J=2$ at
$V=0$ and $V=8$, respectively.
In both cases, curves are fitted with the scaling ansatz
$\chi(L,T)=\chi_0(T)+aL e^{-L/b}$
to extrapolate the spin susceptibility $\chi(T)$ for the infinite system size.
For the data sets with $\beta<6$, we have simulated lattice sizes
$L\times L$ up to $L=30$.


In Fig. \ref{fig:hei_critical}, we present the finite size
scaling of $\chi(T,L)$ in the critical region 
with $\beta$ from 7 to 12.5 and with parameters $V=0$, $J=2$
and $n=1$.
The extrapolated values of $\chi(T)$ are used in Fig. 3 ($a$)
in the main text.
Let us look at the curve with $\beta=10$, the dependence of
$\chi(T,L)$ on $L$ converges at large values of $L$.
The starting of convergence takes places at values of $L$
at the order from 10 to 20, and thus we estimate
the FM correlation length $\xi$ at $\beta=10$
also in this range.

\begin{figure}
\centerline{
\includegraphics[height=0.7\columnwidth,width=0.7\columnwidth]{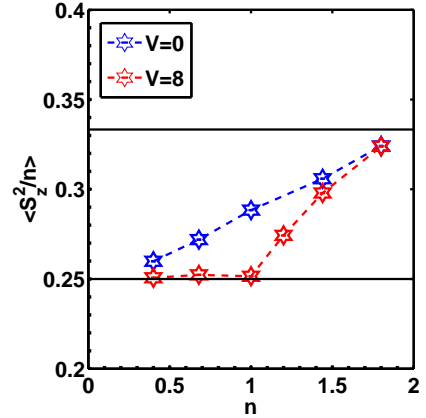}}
\caption{The QMC results for the normalized onsite spin moment $\avg{S_z^2}/n$
at $\beta=6$, $J=2$,  and $L=30$.
Values of $V$ are marked in the inset.
}
\label{fig:sz2}
\end{figure}

In Fig. \ref{fig:sz2}, we plot the QMC simulation results for
the onsite spin moment square normalized by filling density
$\avg{S_z^2}/n$, which behaves
nearly the same as the Curie constants presented in 
Fig. \ref{fig:curie} ($b$).
Due to the SU(2) symmetry, the onsite spin moment square 
$\avg{S^2}/n=3\avg{S_z^2}/n$.
A discussion on $\avg{S_z^2}/n$ is presented in the
Sect. \ref{sect:onsite}.

\begin{figure} [h]
\includegraphics[height=0.49\columnwidth,width=0.49\columnwidth]
{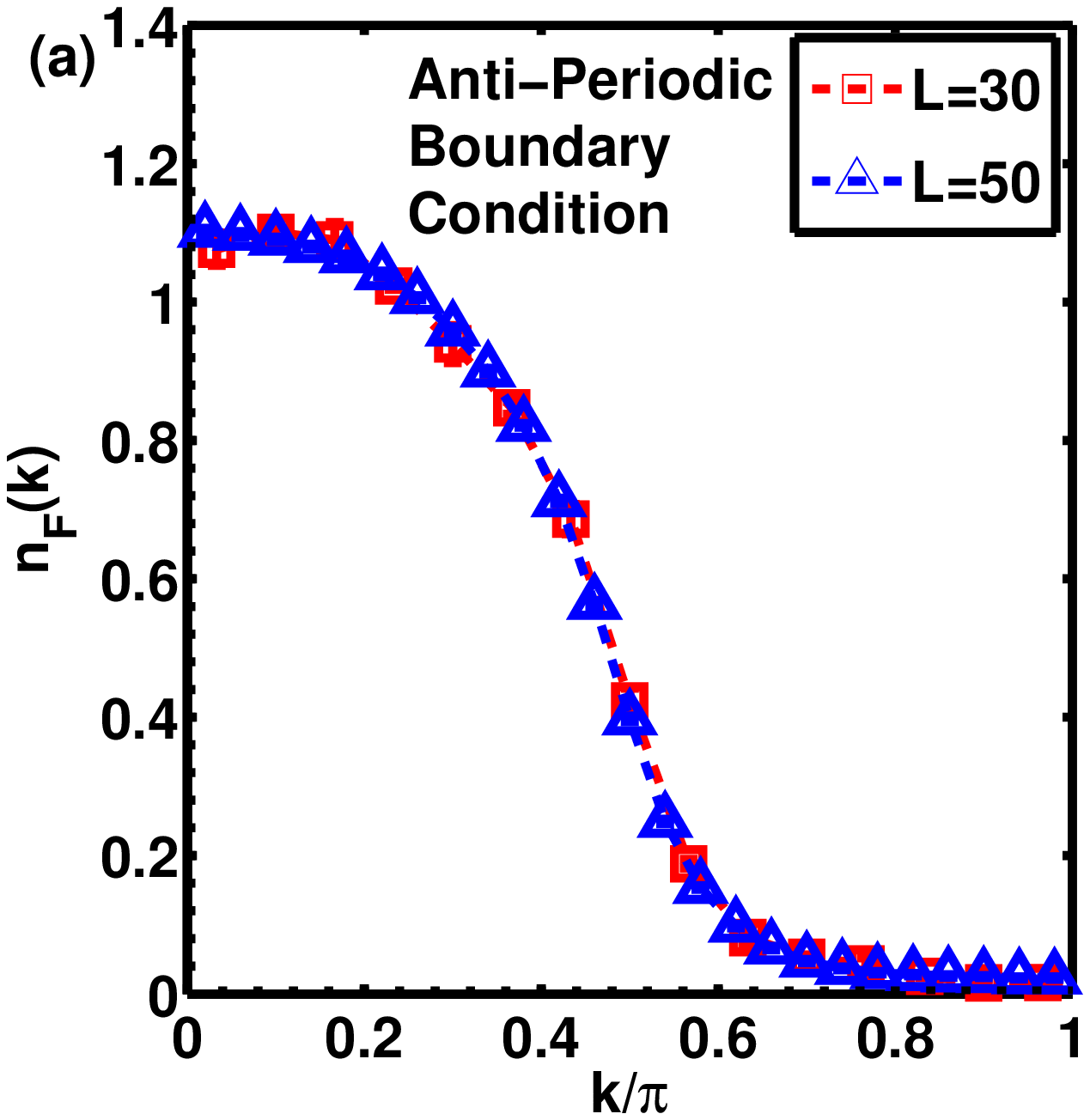}
\includegraphics[height=0.49\columnwidth,width=0.49\columnwidth]
{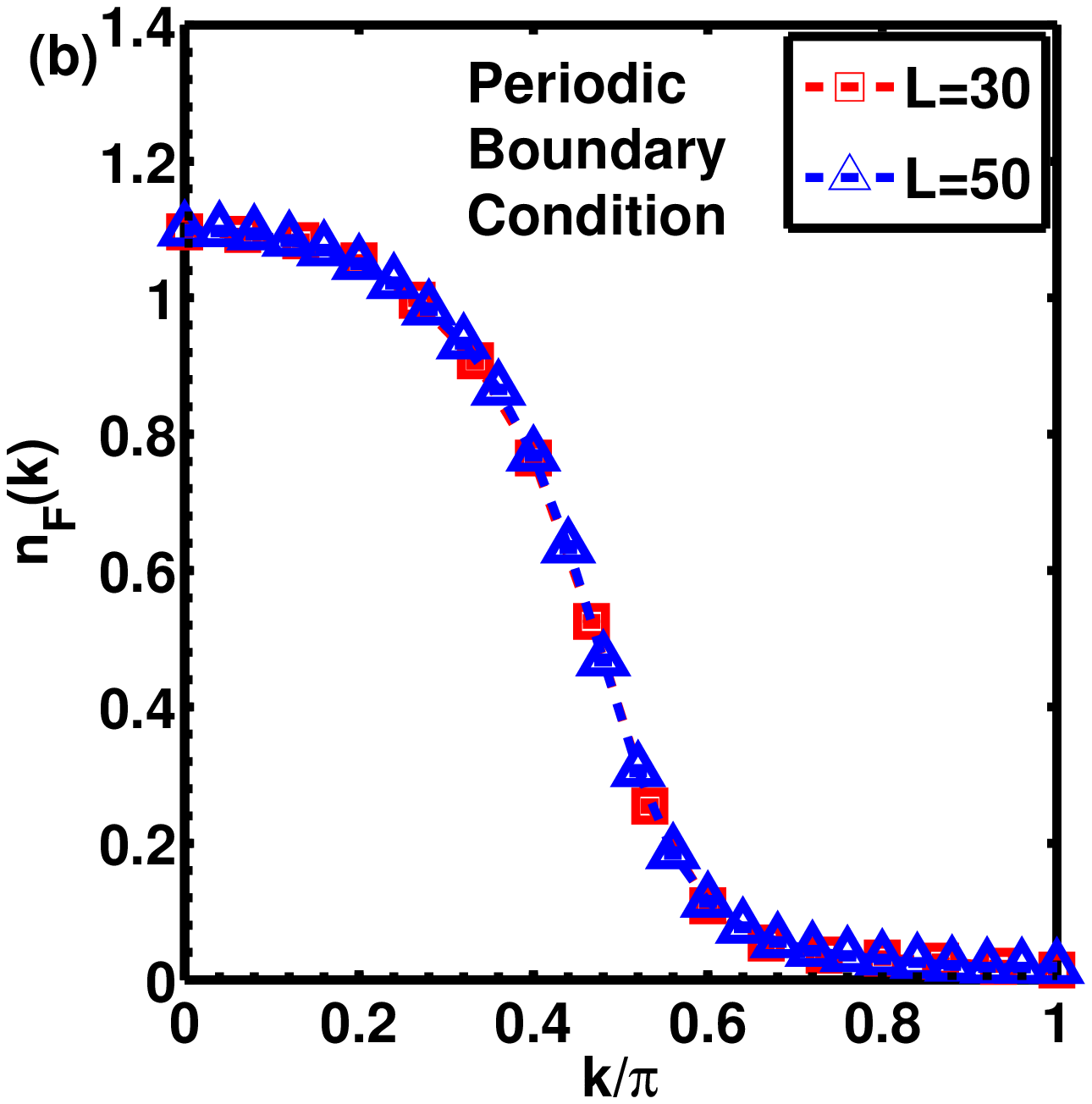}
\caption{The momentum space occupation $n_F(k)$ at $\beta=10$, $V=0$,
$J=2$ and  $n=1$.
Results of $n_F(k)$ for both the anti-periodical boundary condition
and the periodical boundary condition are presented at two
different sizes of $L=30$ and $50$.
}
\label{fig:nk_finite}
\end{figure}

Next we discuss the momentum space distribution $n_F(k)$ which is 1D-like
because the particle number of each chain is separately conserved.
Due to this feature, we make a special choice of boundary
conditions to remove the sign problem: the periodical (anti-periodical)
boundary condition for a chain if its particle number is odd (even).
Our simulation uses the grand canonical assemble, and thus both
configurations with even and odd particle numbers are sampled
which can also be easily distinguished.
For configurations with odd particle numbers, the values of
$k$ take $2p\pi/L$ with $p=0,\pm 1, ... \pm (L-1)$, and for
configurations with even particle numbers, the values of $k$
take $2p\pi/L$ with $p=\pm\frac{1}{2}, ..., \pm (L-\frac{1}{2})$.
Because of this reason, we simulate $n_F(k)$ by separately
sampling configurations of even and odd particle numbers and present
both of them in Fig. \ref{fig:nk_finite} ($a$) and ($b$),
respectively.
Results of the sample sizes with $L=30$ and $50$ are presented,
which shows that the finite size dependence is very weak.
At both sample sizes $L=30$ and $50$, the differences caused by
using periodical or anti-periodical boundary conditions are rather
small.
And the result under the periodical boundary condition is presented
Fig. 1$c$ in the main text.

\section{The orbital ordering at the commensurate filling $n=1$}
\label{append:orbital}

\begin{figure}  
\includegraphics[height=0.8\columnwidth,width=0.7\columnwidth]{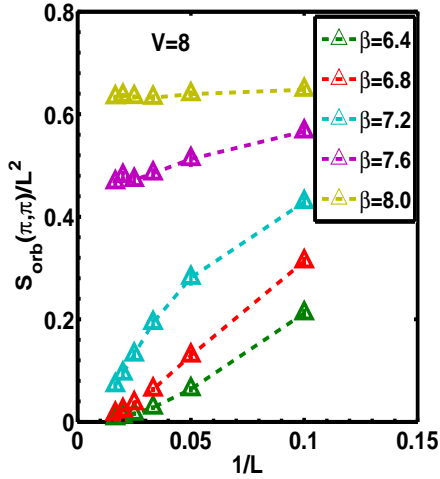}
\caption{The finite size scaling of $S_{orb}(\pi,\pi)/L^2$ with
the parameter values $J_\perp=J_\pp=2$, $n=1$ and $V=8$. 
The system sizes range from $L=10$ to $60$ and the values of 
inverse temperature $\beta$ are presented in the legend.
The orbital ordering occurs at $\beta$ between $7.2$ to $7.6$,
which corresponds to $T_{orb}/t_\pp\approx 0.132\sim 0.139$.
}
\label{fig:orbital}
\end{figure}

Here we present the QMC simulations on the orbital ordering with a 
large value of the inter-orbital repulsion $V$. 
Large $V$ suppresses doubly occupied on-site states, and at the 
commensurate filling $n=1$, the ground state is in the Mott-insulating state.
In this case,  fermions become local moments.
At zero temperature, even though electron spins are fully polarized, 
the orbital degree of freedom enables the superexchange in the 
orbital channel \cite{wu2008}.
The orbital exchange is described by an antiferro-orbital Ising model
\bea
H_{ex}=J_{orb}\sum_{\vec r, \vec r^\prime} \tau_z(\vec r) \tau_z(\vec r^\prime),
\eea
where $J_{orb}=t^2_\pp/V$ and $\tau_z=p^\dagger_x p_x - p^\dagger_y p_y$.
At low temperatures, due to the prominent FM tendency, the above
orbital exchange model still applies, thus below the
temperature scale around $J_{orb}$, the antiferro-orbital
ordering, i.e., the staggered occupation of $p_x$
and $p_y$-orbitals, will appear.

We define the equal-time orbital structure factor as:
\bea
S_{orb}(\vec{q},\tau)=\frac{1}{L^2}\sum\limits_{\vec{r}_1,\vec{r}_2}
\langle m_{orb}(\vec{r}_1,\tau)m_{orb}(\vec{r}_2,\tau)\rangle 
e^{i\vec{q}\cdot(\vec{r}_2-\vec{r}_1)},
\nn \\
\eea
where $m_{orb}(\vec{r},\tau)=n_x(\vec{r}, \tau)-n_y(\vec { r }, \tau)$ is
the on-site orbital polarization. 
Since the orbital ordering occurs at the wavevector $(\pi,\pi)$, we 
present the QMC simulation of the finite scaling of 
$S_{orb}(\pi,\pi)/L^2$ in Fig. \ref{fig:orbital}.
It indicates that the antiferro-orbital ordering appears at
low temperatures, and the critical temperature
$T_{orb}/t_\pp$ lies between $0.132$ and $0.139$.

\begin{figure}
\includegraphics[height=0.6\columnwidth,width=0.6\columnwidth] 
{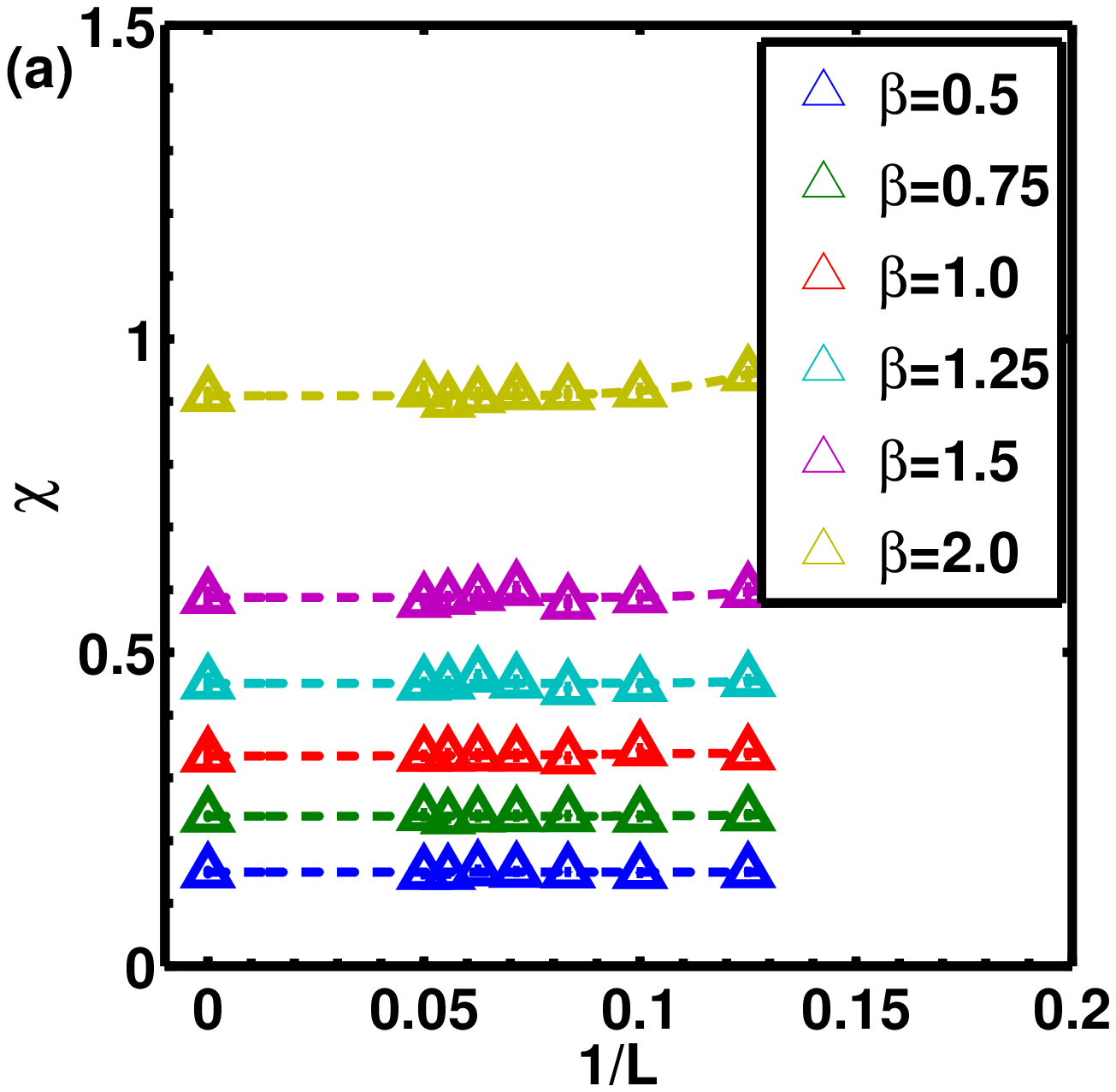}
\includegraphics[height=0.6\columnwidth,width=0.6\columnwidth] 
{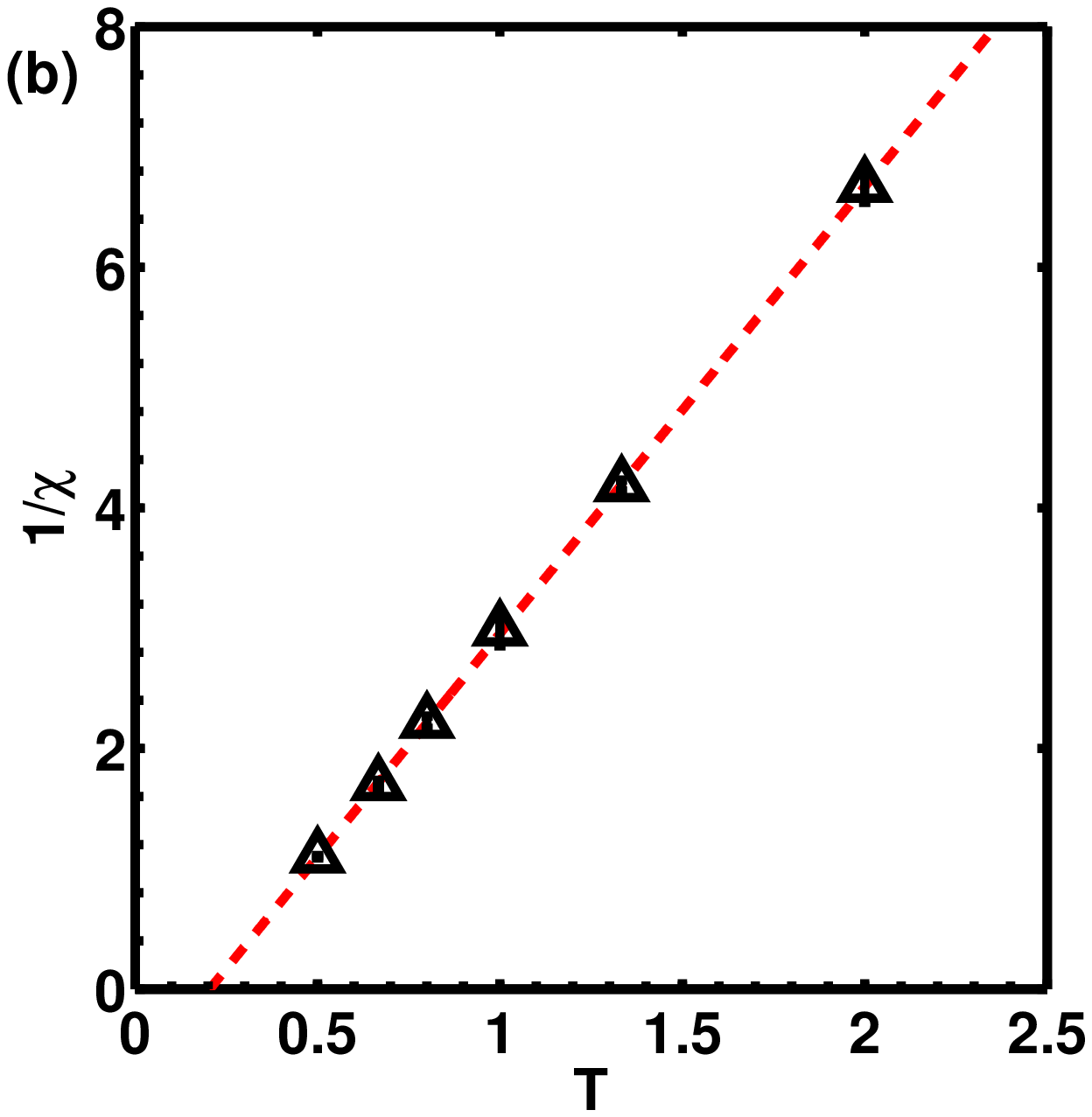}
\caption{(a) The finite size scaling of $\chi(T,L)$ v.s.
$1/L$ in the off-critical region.
The Hund's rule coupling is modified as $J_\perp=2J_\pp=4$.
(b) The CW behavior of the extrapolated $\chi^{-1}(T)$.
The interception on the temperature axis yields
$T_0/t_\pp=0.20\pm 0.01$.
}
\label{fig:ising_h}
\end{figure}

\section{QMC simulations for the Ising class Hamiltonian}
\label{append:ising}

For the Ising class Hamiltonian with the modified Hund's coupling
with $J_\perp=2J_\pp=4$, we extract its mean-field value of the
Curie temperature $T_0$ following the method presented in the main text.
The spin susceptibility $\chi(T)$ is obtained after the finite size
scaling for $\chi(T,L)$ in the off-critical region as shown in
Fig. \ref{fig:ising_h} (a).
By the linear extrapolation of $\chi^{-1}(T)$ in the off-critical region,
we obtain $T_0$ from the interception of $\chi^{-1}(T)$ on the temperature
axis as shown in Fig. \ref{fig:ising_h} (b).
The scaling is performed in the region $T>0.5$ below
which the deviation from the CW behavior appears.
The linear extrapolation of $\chi^{-1}(T)$ gives rise to
$T_{0}=0.20\pm0.01$.
The magnetic structure factors at even lower temperatures in the critical
region are presented
in Fig. \ref{fig:critical_ising} in the main text.



\end{document}